\RequirePackage{fix-cm}
\documentclass[12pt]{iopjournal}
\usepackage{booktabs,tabularx,array,longtable,pdflscape}
\usepackage{float}
\usepackage{placeins}

\usepackage{caption}
\DeclareCaptionFont{figonefont}{\fontsize{8}{9.6}\selectfont}
\usepackage{amsmath,amssymb}
\usepackage[numbers,sort&compress]{natbib}
\hypersetup{hidelinks}
\fancyhfoffset[L]{0mm}
\renewcommand{\articletype}[1]{\noindent{\Large\sffamily Nuclear Fusion}\par\vspace{4mm}\noindent{\scriptsize\sffamily\bfseries\MakeUppercase{#1}}\par\vspace{3mm}}
\begin{document}
\articletype{Paper}
\title{Teacher-guided multi-task learning for integrated plasma monitoring on EAST}
\author{Yunhu Jia\textsuperscript{1,2},
Wenhui Hu\textsuperscript{2,*},
Bihao Guo\textsuperscript{2},
Qingquan Yang\textsuperscript{2},
Xin Lin\textsuperscript{2},
Haoyu Wang\textsuperscript{1,2},
Yao Huang\textsuperscript{2},
Yuehang Wang\textsuperscript{2},
Kai Wu\textsuperscript{2},
Zhengping Luo\textsuperscript{2},
Yanmin Duan\textsuperscript{2},
Jilei Hou\textsuperscript{2},
Ruirui Zhang\textsuperscript{2},
Jianqiu Zhu\textsuperscript{2,*},
Zijie Liu\textsuperscript{3},
Qiping Yuan\textsuperscript{2},
Bingjia Xiao\textsuperscript{3,2,1}}
\par
\affil{$^1$University of Science and Technology of China, Hefei 230026, Anhui, China}
\par
\affil{$^2$Institute of Plasma Physics, Hefei Institutes of Physical Science (HFIPS), Chinese Academy of Sciences, Hefei 230026, Anhui, China}
\par
\affil{$^3$Institute of Energy, Hefei Comprehensive National Science Center, Hefei 230001, Anhui, China}
\par
\affil{$^*$Corresponding authors.}
\par
\email{\href{mailto:huwh@ipp.ac.cn}{huwh@ipp.ac.cn}, \href{mailto:jqzhu@ipp.ac.cn}{jqzhu@ipp.ac.cn}}
\par
\begin{abstract}
This paper integrates tokamak plasma disruption prediction, Edge-Localized Mode (ELM) monitoring, Multifaceted Asymmetric Radiation From the Edge (MARFE) detection, and H/L-mode identification into a unified Multi-Task Learning (MTL) model.
This integration aims to realize MTL's potential for improving task performance while reducing inference time.
However, on datasets constructed from selected EAST discharges recorded between 2015 and 2020, this study found that the traditional MTL method underperformed the corresponding Single-Task Learning (STL) method on some tasks.
This performance degradation is the well-known issue of negative transfer.
To address this issue, this study proposes the teacher-guided MTL method, using task-specific STL ensembles (teachers) to guide the training of an MTL model (student).
The ensemble teachers guide training through soft-target distillation losses for each task of the MTL student model, and increased weighting of each shot's peak-disruptivity window in the disruption prediction label-supervised loss.
Test-set results showed that the teacher-guided MTL method outperformed the corresponding STL method in ROC-AUC for all four tasks, eliminating the negative transfer observed with the traditional MTL method.
In addition, on the same single GPU or single physical CPU core, the MTL student model can substantially reduce the inference time required to generate all four task outputs compared with the four STL baseline models.
This paper offers a promising route to plasma monitoring systems that cover a broader range of phenomena, improve per-task performance and reduce inference time with limited real-time computational resources, supporting future high-performance plasma operation.
\end{abstract}

\section{Introduction}
\label{sec:introduction}

Automatic monitoring and prediction of plasma phenomena are important for high-performance plasma operation in tokamaks.
This study focuses on four plasma monitoring tasks: Disruption Prediction (DP), Multifaceted Asymmetric Radiation from the Edge (MARFE) detection, Edge-Localized Mode (ELM) detection, and High/Low-Confinement-Mode (H/L) identification.
The high stored energy and plasma current of high-performance plasmas can make disruptions particularly damaging to the device, requiring reliable DP to trigger protective measures in time~\cite{hender2007mhd,de2016requirements}.
High-performance plasma operation typically relies on sustained H-mode confinement, motivating automatic H/L-mode identification to track the confinement state~\citep{doyle2007confinement}.
ELMs during H-mode operation impose transient heat and particle loads that can damage plasma-facing components, motivating ELM detection for control~\cite{bandyopadhyay2025mhd}.
Higher plasma density is desirable for increasing fusion power, but the density limit must be avoided~\citep{Bernert2015}.
MARFE onset can indicate proximity to this limit~\citep{giacomin2022density}, motivating automatic MARFE detection~\citep{GonzalezGanzabal2024}.
Together, these four tasks provide comprehensive information on plasma phenomena to support targeted intervention in future high-performance plasma operation.

The need for more accurate monitoring results motivates the widespread use of data-driven methods~\citep{kates2019predicting,guo2021fullcnn,guo2023disruption,seo2024avoiding,seo2023multimodal,Transfer2023,orozco2022neural,yang2023neural,Zhu2023,GonzalezGanzabal2024,Deng2025}.
A rapidly developing trend is to use Multi-Task Learning (MTL) to integrate tasks into one model to improve task performance.
Ref.~\citep{Zhu2023} combined unstable-event identification with DP, improving DP's shot-by-shot ROC-AUC and median warning lead time over a disruption-only model.
Ref.~\citep{Deng2025} integrated H/L-mode identification and ELM detection, achieving 96.7\% overall classification accuracy, higher than that of Single-Task Learning (STL) models.
Ref.~\citep{ZhuXB2025} jointly detected three Magnetohydrodynamic (MHD) instabilities (fishbone, long-lived, and tearing modes) on the HL-2A tokamak, achieving higher window-level ROC-AUC than STL for each.

Another motivation for MTL is to reduce inference time when computational resources are limited.
Plasma phenomena are diverse.
For example, a survey of disruption causes on the JET tokamak identified 20 types of related physical phenomena~\citep{devries2011survey}.
Deploying a separate STL model per task therefore adds computational load as monitoring expands.
EAST's Plasma Control System (PCS) main control cabinet has 16 CPU cores and two GPU cards, whose existing workload limits capacity for additional monitoring~\citep{yuan2018pcsupgrade,huang2022realtimecontrol}.
This added load may push inference time beyond the budget for control updates or leave less time to implement protective measures against disruptions.
MTL can reduce inference time by sharing feature extraction across tasks, so that shared features are computed once and reused for multiple task outputs.

However, MTL does not consistently improve task performance and may underperform the corresponding STL models on some tasks, an issue known as negative transfer~\cite{ruder2017overviewmultitasklearningdeep}.
Whether negative transfer occurs and how difficult it is to mitigate can depend on the task combination.
The above H/L-mode identification~\citep{Deng2025} and MHD detection~\citep{ZhuXB2025} studies did not report negative transfer with task-label losses alone.
In contrast, previous studies on equilibrium reconstruction~\citep{sun2024impact} and pedestal prediction~\citep{gillgren2022enabling} reported that some jointly learned outputs had lower \(R^2\) values than those obtained with the corresponding STL models.
Ref.~\citep{chen2026fusionmmt} integrated DP, ELM recognition, and regression of the energy-confinement enhancement factor H98 on EAST.
To mitigate negative transfer, Ref.~\citep{chen2026fusionmmt} used TaskAdapt modules, task-specific gates, and three-stage Progressive Multitask Optimization with uncertainty weighting and task-specific refinement.
Despite these measures, DP AUC in Ref.~\citep{chen2026fusionmmt} remained below that of the corresponding STL model.
Negative transfer therefore remains a major obstacle to integrating more plasma monitoring tasks into MTL without degrading task performance.

This study integrates DP, MARFE detection, ELM detection, and H/L-mode identification into an MTL model, using datasets compiled from selected EAST discharges recorded during the 2015--2020 experimental campaigns.
The traditional MTL method, which uses only task-label supervision, exhibits negative transfer for this task combination.
To mitigate negative transfer, this paper introduces the teacher-guided MTL method.
This method uses independently optimized task-specific STL ensembles as teachers to guide a unified MTL student model through soft-target distillation across all four tasks.
The DP teacher also identifies each training shot's peak-disruptivity window, which receives greater weight in the student's DP label-supervised loss.
After training, the MTL student model runs without the teachers, preserving the efficiency of shared computation.

To our knowledge, no previous study has used a distillation loss to mitigate negative transfer in tokamak plasma multi-task monitoring.
Nor has any study proposed a teacher-guided strategy for identifying each shot's peak-disruptivity window and increasing its weight in the label-supervised loss.
Regarding distillation, Ref.~\citep{dong2025pst} used it to compensate for accuracy loss in a lightweight plasma shape detection model, supporting faster inference, but did not evaluate its role in mitigating negative transfer.
In other fields, such as natural language processing, soft-target distillation has been used to improve MTL performance~\citep{liu2019improving,meng2021multitaskdistillation}.
However, tokamak plasma monitoring has domain-specific characteristics: for example, different tasks may require different input signals, and the physical phenomena involved may have different characteristic time scales.
Results from other fields therefore do not directly demonstrate that distillation is also effective in plasma monitoring.
For teacher-guided disruption prediction, only Ref.~\citep{cai2024improved} proposed a label-correction strategy based on the teacher's classification results.
In contrast, our DP peak-disruptivity window-weighting strategy does not correct labels but uses teacher guidance to make the student model focus more on peak-disruptivity windows during training.

The remainder of this paper is organized as follows.
Section~\ref{subsec:database-task-setup} describes the datasets, and Section~\ref{subsec:model-architecture-evaluation} presents the model architecture and explains how task evaluation metrics are calculated from model outputs.
Section~\ref{subsec:baseline-construction} describes the construction of the four STL baseline models and the traditional MTL baseline model, while Section~\ref{subsec:multi-task-ensemble-distillation} details the teacher-guided MTL method.
Section~\ref{subsec:results-hpo-validation} reports the results for the STL baseline models and task-specific ensemble teachers.
Sections~\ref{subsec:results-negative-transfer}--\ref{subsec:results-efficiency} evaluate the mitigation of negative transfer, component contributions through ablation experiments, and the reduction in inference time, respectively.
Section~\ref{sec:conclusions} summarizes the findings and outlines future work.

\section{Methods}
\label{sec:methods}
Using datasets for the four EAST monitoring tasks, this study evaluates whether the teacher-guided MTL method improves per-task performance and reduces inference time.
Four STL baseline models provide references for per-task performance and the total inference time required to generate all four task outputs on a single GPU or a single physical CPU core.
The traditional MTL baseline model, trained without teacher guidance, is compared with the corresponding STL baseline models to assess negative transfer.

\subsection{Database}
\label{subsec:database-task-setup}
The four task datasets contain data collected during EAST experimental campaigns between 2015 and 2020 and collectively involve 40 diagnostic and operational signals from EAST.
These signals are divided into 15 signal groups according to the physical quantities they represent: 19 XUV radiation signals, four deuterium Balmer-alpha emission signals, five heating-power signals (electron cyclotron resonance heating, $P_{\mathrm{ECRH}}$; lower hybrid wave heating and current drive, $P_{\mathrm{LHW}}$; ohmic heating, $P_{\mathrm{OHM}}$; neutral beam injection heating, $P_{\mathrm{NBI}}$; and heating in the ion cyclotron range of frequencies, $P_{\mathrm{ICRF}}$), and 12 groups containing one signal each.
These 40 signals form the combined input signal inventory for the four tasks. The task-specific signal-group selection rules are described in Section~\ref{subsec:model-architecture-evaluation}.
The signal groups and the number of signals in each group are summarized in Table~\ref{tab:input-signal-groups}.
All signals are interpolated to 1\,kHz, and signals within the same group use the same normalization parameters.

\begingroup
\small
\renewcommand{\arraystretch}{1.15}
\setlength{\tabcolsep}{4pt}
\newlength{\SignalDescriptionWidth}
\settowidth{\SignalDescriptionWidth}{\begin{tabular}{@{}clc@{}}Group No. & $I_{p,\mathrm{error,norm}}$ & Signal count\end{tabular}}
\setlength{\SignalDescriptionWidth}{\dimexpr\linewidth-\SignalDescriptionWidth-2\tabcolsep\relax}
\setlength{\LTcapwidth}{\linewidth}
\setcounter{LTchunksize}{200}
\setlength{\LTpre}{4pt}\setlength{\LTpost}{4pt}
\begin{longtable}{@{}cl>{\raggedright\arraybackslash}p{\SignalDescriptionWidth}c@{}}
\caption{Signal groups used across the four tokamak plasma monitoring tasks.}
\label{tab:input-signal-groups}\\
\toprule
Group No. & Symbol & Description & Signal count \\
\midrule
\endfirsthead
\multicolumn{4}{@{}l@{}}{\tablename~\thetable\ (continued)}\\[3pt]
\toprule
Group No. & Symbol & Description & Signal count \\
\midrule
\endhead
\midrule
\multicolumn{4}{r}{\footnotesize Continued on next page}\\
\endfoot
\bottomrule
\noalign{\vskip4pt}
\multicolumn{4}{@{}l@{}}{\begin{minipage}{\linewidth}
\footnotesize\raggedright
\textit{Note.} Signal-group selection differs among tasks.
\end{minipage}}\\
\noalign{\vskip0pt}
\endlastfoot
1 & $PXUV$ & XUV radiation & 19 \\
2 & $D_\alpha$ & Deuterium Balmer-alpha emission & 4 \\
3 & $P_{\mathrm{heat}}$ & Heating powers: $P_{\mathrm{ECRH}}$, $P_{\mathrm{LHW}}$, $P_{\mathrm{OHM}}$, $P_{\mathrm{NBI}}$, $P_{\mathrm{ICRF}}$ & 5 \\
4 & $P_{\mathrm{RAD}}$ & Total radiated power & 1 \\
5 & $V_{\mathrm{loop}}$ & Loop voltage & 1 \\
6 & $Z_{\mathrm{error}}$ & Fast vertical-control error (high-pass filtered) & 1 \\
7 & $I_p$ & Plasma current & 1 \\
8 & $I_{p,\mathrm{error,norm}}$ & Normalized current tracking error, $I_{p,\mathrm{error}}/I_p$ & 1 \\
9 & $f_{\mathrm{GW}}$ & Greenwald density fraction, $\overline{n}_e\pi a^2/I_p$ & 1 \\
10 & $\kappa$ & Plasma elongation & 1 \\
11 & $l_i$ & Internal inductance & 1 \\
12 & $A_{n=1}$ & Saddle-coil $n=1$ mode amplitude & 1 \\
13 & $q_{95}$ & Safety factor at the 95 percent flux surface & 1 \\
14 & $a_{minor}$ & Plasma minor radius & 1 \\
15 & $W_{mhd}$ & Plasma stored energy & 1 \\
\end{longtable}
\endgroup

For each task, a domain expert familiar with the corresponding plasma phenomenon provides the annotations.
The ITER Research Plan (2018)~\cite{iter2018researchplan} describes distinct plasma evolution and control objectives for the ramp-up, flat-top, and ramp-down phases.
To avoid the additional modeling and evaluation complexity arising from these phase differences, this study follows previous DP studies~\cite{guo2023disruption,Transfer2023,rea2019realtime,montes2019machine} and restricts both data annotation and the models' operating scope to the flat-top phase.

These flat-top discharge data are partitioned into training, validation, and test sets by shot, with all samples from the same shot assigned to the same subset across all tasks.
The training set is used to optimize model parameters, the validation set for hyperparameter optimization and model selection, and the test set only for independent final evaluation.
All four tasks use sliding windows of 100\,ms to construct model inputs, with window labels and strides determined by the task and the purpose of each data subset.
Table~\ref{tab:dataset-partitioning} summarizes the shot counts and positive and negative window counts for each task and data subset.

For ELM detection, MARFE detection, and H/L-mode identification, both training and evaluation use window-level labels.
For ELM detection, windows containing an ELM peak are labelled positive.
For MARFE detection, a window is labelled positive if MARFE is present at its end time.
For H/L-mode identification, a window is labelled positive if its end time falls within an H-mode interval.
Windows that do not meet the corresponding condition are labelled negative.
Given limited computational resources, the training and validation sets for these three tasks use a stride of 10\,ms to reduce the time required for model training and hyperparameter optimization, whereas the test sets use a stride of 1\,ms for final evaluation.

For DP, each discharge is first labelled disruptive if a disruption occurs during the flat-top phase and non-disruptive otherwise.
These shot-level labels are used for validation and test evaluation, in which window-level predictions are generated with a stride of 1\,ms and then used to calculate shot-level performance metrics. Window-level labels are therefore not required.
The detailed evaluation protocol is described in Section~\ref{subsec:model-architecture-evaluation}.
The DP training set requires converting the shot-level labels into window-level supervision.
For disruptive discharges, windows whose end times fall within the final 200\,ms before the current quench are labelled positive, while earlier windows are excluded from training; windows from non-disruptive discharges are labelled negative.
The 200\,ms boundary follows Guo et al.~\cite{guo2021fullcnn,guo2023disruption}; Ref.~\citep{guo2021fullcnn} selected it as a representative starting point for disruption evolution based on statistical analysis of EAST diagnostic signals.
Although precursor onset varies among discharges, EAST DP models using this fixed labeling boundary have achieved effective prediction~\cite{guo2021fullcnn,guo2023disruption}.
This boundary is therefore adopted to construct training labels and reduce the annotation effort required for experts to identify the precise onset of disruption precursors in each discharge.
Under these labeling rules, DP training windows are generated with a stride of 10\,ms for disruptive discharges and 100\,ms for non-disruptive discharges to mitigate the severe imbalance between positive and negative training samples.

\begingroup
\small
\renewcommand{\arraystretch}{1.2}
\setlength{\tabcolsep}{4pt}
\setlength{\LTcapwidth}{\linewidth}
\setcounter{LTchunksize}{200}
\setlength{\LTpre}{4pt}\setlength{\LTpost}{4pt}
\begin{longtable}{@{}>{\raggedright\arraybackslash}p{0.10\linewidth}>{\raggedright\arraybackslash}p{0.24\linewidth}*{3}{>{\centering\arraybackslash}p{\dimexpr(0.66\linewidth-8\tabcolsep)/3\relax}}@{}}
\caption{Details of dataset annotation and partitioning.}
\label{tab:dataset-partitioning}\\
\toprule
Task & Count type & Training & Validation & Test \\
\midrule
\endfirsthead
\multicolumn{5}{@{}l@{}}{\tablename~\thetable\ (continued)}\\[3pt]
\toprule
Task & Count type & Training & Validation & Test \\
\midrule
\endhead
\midrule
\multicolumn{5}{r}{\footnotesize Continued on next page}\\
\endfoot
\bottomrule
\noalign{\vskip4pt}
\multicolumn{5}{@{}l@{}}{\begin{minipage}{\linewidth}
\footnotesize\raggedright
\textit{Note.} Dis./Non-Dis.: disruptive/non-disruptive shots; Pos./Neg.: positive/negative windows.
N/A: no window-level labels for DP validation or testing.
Each shot is assigned to the same data subset (training, validation, or test) across all tasks in which it appears.
\end{minipage}}\\
\noalign{\vskip0pt}
\endlastfoot
DP & Shots (Dis./Non-Dis.) & 604\,/\,1267 & 511\,/\,1189 & 488\,/\,1174 \\*
 & Windows (Pos./Neg.) & 12080\,/\,93677 & N/A & N/A \\
\addlinespace[0.5em]
ELM & Shots & 150 & 185 & 122 \\*
 & Windows (Pos./Neg.) & 18129\,/\,27815 & 16545\,/\,21076 & 151560\,/\,242480 \\
\addlinespace[0.5em]
MARFE & Shots & 79 & 103 & 55 \\*
 & Windows (Pos./Neg.) & 4591\,/\,23914 & 2722\,/\,17658 & 17150\,/\,95550 \\
\addlinespace[0.5em]
H/L & Shots & 92 & 87 & 82 \\*
 & Windows (Pos./Neg.) & 10878\,/\,10844 & 7692\,/\,9190 & 119050\,/\,115260 \\
\end{longtable}
\endgroup

\setlength{\intextsep}{8pt plus 2pt minus 2pt}
\subsection{Model architecture and task performance evaluation}
\label{subsec:model-architecture-evaluation}
\begin{figure}[!t]
\captionsetup{font=figonefont}
\centering
\includegraphics[width=\linewidth]{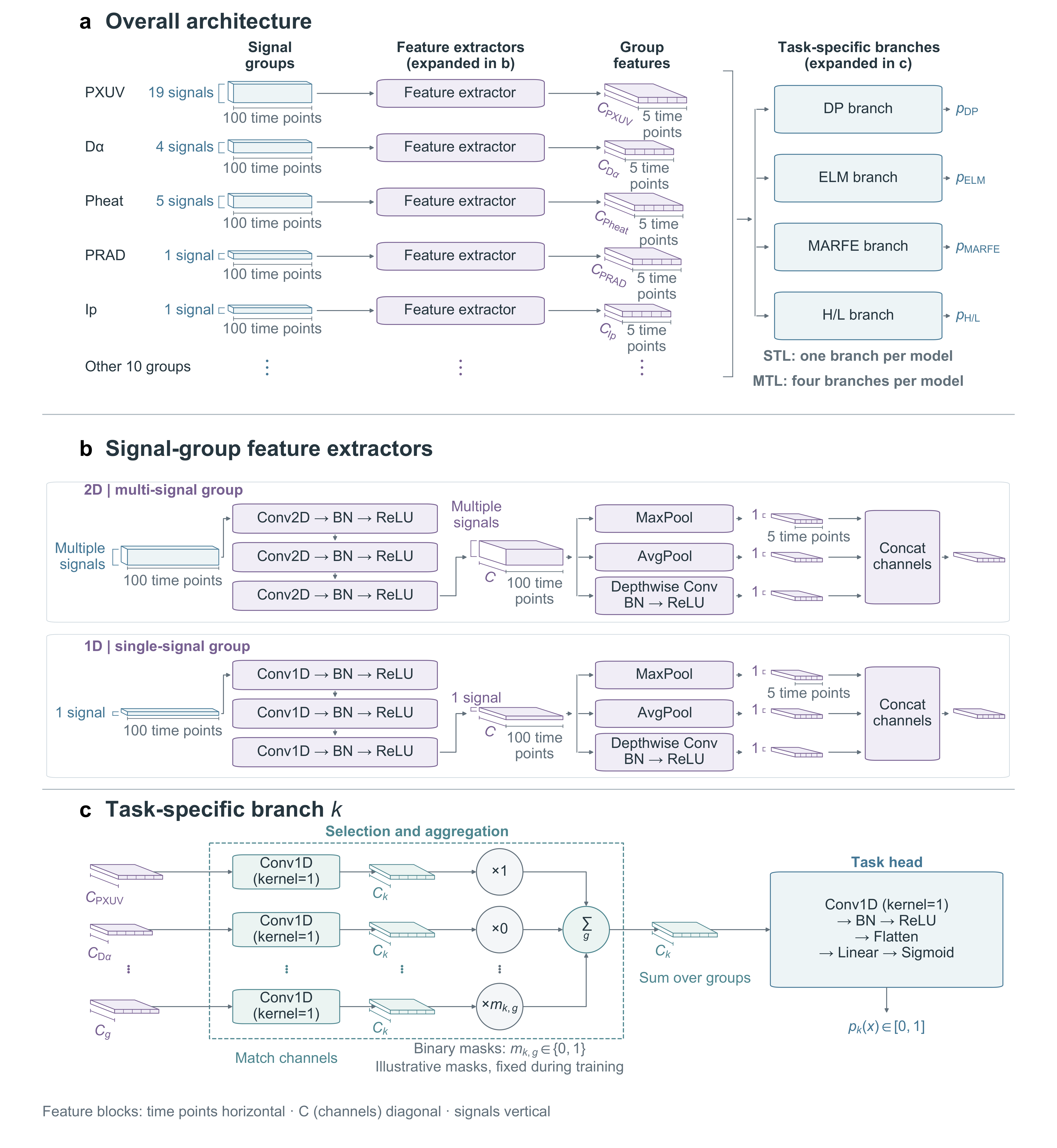}
\caption{Model architecture.
(a) Overall architecture: each STL model uses one task-specific branch, whereas each MTL model uses four.
(b) Signal-group feature extractor.
(c) Task-specific branch.
Feature blocks show time horizontally, channels diagonally, and signal count vertically.
Binary masks are illustrative, and dimensions are not drawn to scale.}
\label{fig:model-architecture}
\end{figure}

This study uses STL and MTL models with the modular architecture shown in Figure~\ref{fig:model-architecture}a, consisting of signal-group feature extractors and task-specific branches.
Each STL model has one task-specific branch, whereas each MTL model has four.

\par\ifdim\dimexpr\pagegoal-\pagetotal\relax<3\baselineskip\newpage\fi
\looseness=-1
Each signal group has its own feature extractor, with parameters independent of those for other groups, to process its 100-ms input windows (Figure~\ref{fig:model-architecture}b).
Each extractor contains three modules comprising convolution, Batch Normalization (BN), and a Rectified Linear Unit (ReLU), with one-dimensional convolutions for single-signal groups and two-dimensional convolutions for multi-signal groups.
After these modules, maximum pooling, average pooling, and depthwise convolution operate in parallel to reduce the temporal dimension from 100 to five positions and, for multi-signal groups, the signal dimension to one.
The outputs of the three layers are concatenated along the channel dimension to form the signal-group feature map $\mathbf{F}_g$.

The task-specific branch for task $k$ converts signal-group features into the task output through group-wise projection, signal-group selection, feature aggregation, and a task head (Figure~\ref{fig:model-architecture}c).
Each group feature map $\mathbf{F}_g$ is projected to a common channel count $C_k$ by a group-specific one-dimensional convolution $P_{k,g}$ with kernel size 1.
The signal groups used by task $k$ are determined by its binary selection-mask hyperparameters $m_{k,g}\in\{0,1\}$: a value of 1 retains the group, whereas 0 excludes it.
For ELM, MARFE, and H/L, these signal-group selection masks are jointly optimized with architecture and training hyperparameters during hyperparameter optimization, guided by validation performance to suit each task's diagnostic requirements and reduce irrelevant or redundant inputs (see Section~\ref{subsec:baseline-construction}).
For DP, all 15 groups and their 40 signals are retained, with $m_{\mathrm{DP},g}=1$, because disruptions may involve a wide range of plasma phenomena and a larger dataset is available for this task.
\par\begin{samepage}
The selected projected features are then summed to form the task feature map $\mathbf{Y}_k$:
\begin{equation}
\mathbf{Y}_k=\sum_{g=1}^{15}m_{k,g}P_{k,g}(\mathbf{F}_g),
\qquad \mathbf{Y}_k\in\mathbb{R}^{C_k\times5}.
\label{eq:baseline-group-aggregation}
\end{equation}
\end{samepage}
The task head processes the aggregated feature map $\mathbf{Y}_k$ through a one-dimensional convolution with kernel size 1, BN, ReLU, flattening, and a linear layer, followed by a sigmoid to produce the task output probability $p_k(x)$ for input window $x$.

\begin{figure}[!t]
\centering
\captionsetup{font=figonefont}
\includegraphics[width=\linewidth]{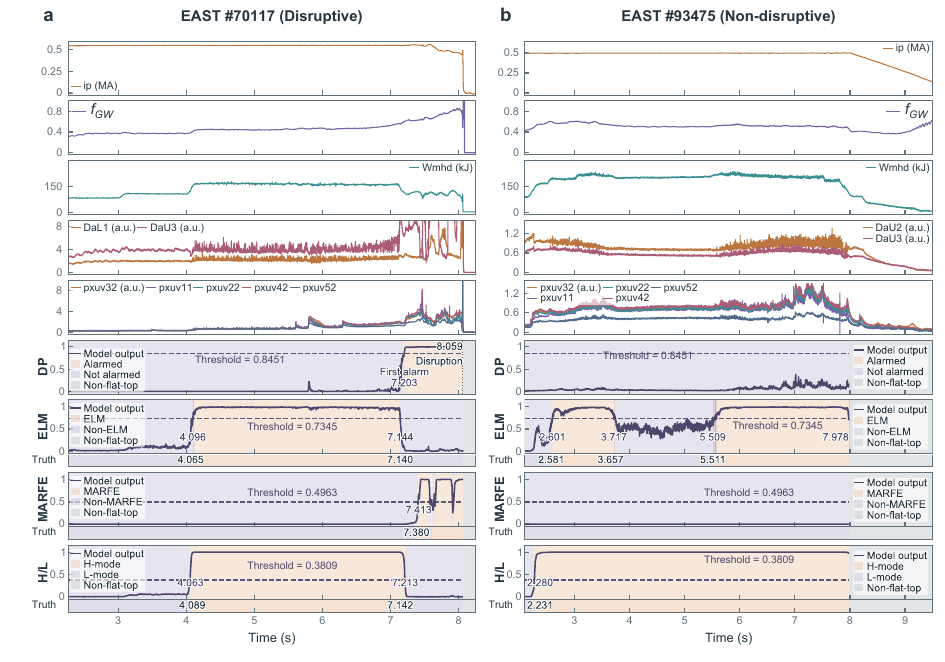}
\caption{Illustrative input signals and four-task outputs for (a) disruptive EAST shot \#70117 and (b) non-disruptive EAST shot \#93475. The upper panels show a subset of the 40 input signals. The Truth bands show expert-annotated ground truth.}
\label{fig:input-output-example}
\label{fig:input-output-example-93475}
\end{figure}

Figure~\ref{fig:input-output-example} illustrates selected input signals and the four task output probability curves $p_k(x)$ for a disruptive EAST shot \#70117 and a non-disruptive shot \#93475.
These probability curves are converted into classification results by applying a decision threshold, with the evaluation unit determined by the task.
For ELM detection, MARFE detection, and H/L-mode identification, each window is classified as positive if its output probability exceeds the threshold and as negative otherwise.
Comparison with the task-specific window labels gives True Positives (TP) and False Negatives (FN) for positively labelled windows classified as positive and negative, respectively, and True Negatives (TN) and False Positives (FP) for negatively labelled windows classified as negative and positive, respectively.
For DP, the probability curve is instead used to determine whether and when an alarm occurs during each shot, and classification is evaluated at the shot level.
For each shot, the alarm time is defined as the first time the predicted probability strictly exceeds the threshold.
A disruptive shot is counted as a TP if the alarm occurs at least 10\,ms before disruption. Other alarms with less than 10\,ms lead time, or no alarm, give an FN.
A non-disruptive shot is counted as an FP if an alarm occurs, and as a TN otherwise.

\begin{samepage}
At each threshold, TP, TN, FN, and FP are aggregated over the evaluation dataset to calculate the true-positive rate, $\mathrm{TP}/(\mathrm{TP}+\mathrm{FN})$, and false-positive rate, $\mathrm{FP}/(\mathrm{FP}+\mathrm{TN})$; sweeping the threshold yields the ROC curve.
The area under this curve is termed WindowAUC for the window-level evaluation of ELM, MARFE, and H/L, and ShotAUC for the above shot-classification protocol for DP.
On the independent test set, an MTL model exhibits negative transfer if its AUC for any task is lower than that of the corresponding STL baseline model.

\par\end{samepage}

\FloatBarrier
\subsection{Baseline construction}
\label{subsec:baseline-construction}
\label{subsec:single-task-baselines}
\begin{figure}[!htbp]
\centering
\captionsetup{font=figonefont}
\includegraphics[width=\linewidth]{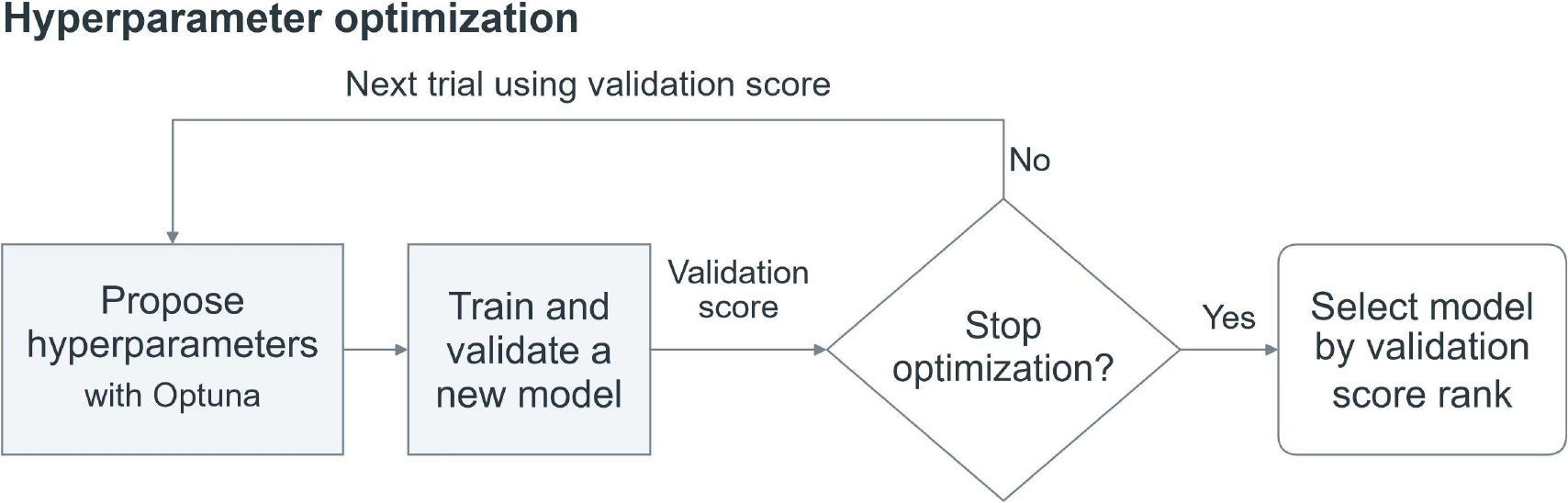}
\caption{Model training and hyperparameter optimization.
Each trial trains and validates a new model.
Models are ranked by validation score for selection.
Training and checkpoint selection follow Section~\ref{subsec:baseline-construction}.
}
\label{fig:training-hyperparameter-optimization}
\end{figure}

To assess negative transfer, four STL baseline models and one traditional MTL baseline model are selected through separate hyperparameter optimization processes.
In each trial, a model is trained and validated using a fixed set of hyperparameters.
After the initial random trials, Optuna uses the Tree-Structured Parzen Estimator (TPE) to propose hyperparameters for the next trial based on previous trials' validation scores.
Models are ranked by validation score for selection.
The overall procedure is shown in Figure~\ref{fig:training-hyperparameter-optimization}.

\FloatBarrier

Hyperparameter optimization covers model architecture, training, and signal-group selection.
The architecture hyperparameters include feature-extractor channel counts and convolution kernel sizes, the projection channel count $C_k$, and the task-head hidden width $H_k$.
The training hyperparameters include the AdamW learning rate and weight-decay setting for both STL and MTL; the task-loss weights $\lambda_k$ are used only in MTL to control the contributions of the four task-label losses to the total training loss.
The signal-group selection hyperparameters are the masks introduced in Section~\ref{subsec:model-architecture-evaluation}.
For ELM, MARFE, and H/L, these masks are optimized jointly with architecture and training hyperparameters during hyperparameter optimization for the STL and traditional MTL methods.
DP uses all signal groups because disruptions may involve a wide range of plasma phenomena and a larger dataset is available for this task.
The hyperparameters and their applicability are listed in Table~\ref{tab:baseline-configuration} in Appendix~\ref{app:baseline-configuration}.
Hyperparameter values remain fixed throughout model training.

Within each trial, the model is trained using one random seed and the AdamW optimizer.
Each STL model minimizes the Binary Cross-Entropy (BCE) task-label loss for its corresponding task:
\begin{equation}
\mathcal{L}_k=-\frac{1}{N_k}\sum_{i=1}^{N_k}
\left[y_{k,i}\log p_k(x_{k,i})+
(1-y_{k,i})\log(1-p_k(x_{k,i}))\right],
\label{eq:baseline-task-loss}
\end{equation}
where $p_k(x_{k,i})$ is the predicted positive-class probability for input window $x_{k,i}$ of task $k$, $y_{k,i}\in\{0,1\}$ is its label, and $N_k$ is the number of windows with valid labels for that task.
Unlabelled windows are excluded from both the loss sum and its normalization.
At the end of each epoch, validation ShotAUC for DP or WindowAUC for ELM, MARFE, and H/L is used for checkpoint selection and to determine whether early stopping is required.
If the corresponding validation metric does not improve for ten consecutive epochs, model training stops and the checkpoint with the highest validation AUC is restored.

Traditional MTL models undergo two-stage training: the first stage trains the entire network, and the second freezes the shared feature extractors and refines the task-specific branches separately.
In the first stage, the four task-label losses defined above are combined using the task-loss weights to form the training objective for the entire network:
\begin{equation}
\mathcal{L}_{\mathrm{MTL}}=\sum_{k=1}^{4}\lambda_k\mathcal{L}_k,
\label{eq:baseline-mtl-loss}
\end{equation}
where $\lambda_k$ is the task-loss weight specified for that trial.

During hyperparameter optimization, STL models are compared using their task's validation AUC, whereas traditional MTL models are compared using validation performance across all four tasks and the inference time required to generate all four outputs.
The MTL composite validation score, $\mathrm{ValScore}_{\mathrm{MTL}}$, is therefore defined for first-stage checkpoint selection and early stopping, and for comparing the trained models across trials.
The score combines overall performance across the four tasks relative to their STL baselines, performance gains beyond those baselines, and inference time:
\begin{equation}
\begin{aligned}
\mathrm{ValScore}_{\mathrm{MTL}}
&=\left(\prod_{k=1}^{4}\operatorname{clip}(\mathrm{AUC}_{k,\mathrm{norm}},0,1)\right)^{1/4}\\
&\quad+0.0001\sum_{k=1}^{4}\max\!\left(0,\mathrm{AUC}_{k,\mathrm{norm}}-1\right)\\
&\quad+0.0001\,\mathrm{InfTimeScore}.
\end{aligned}
\label{eq:performance-efficiency}
\end{equation}
Here, $\mathrm{AUC}_{k,\mathrm{norm}}$ is the validation AUC for task $k$ normalized against fixed STL reference values, $\mathrm{InfTimeScore}$ is the inference-time score, and $\operatorname{clip}(\cdot,0,1)$ limits a value to $[0,1]$.

The first term evaluates performance across all four tasks relative to their STL baselines through the geometric mean of the clipped normalized validation AUCs.
Each task's validation AUC is normalized using fixed STL reference values:
\begin{equation}
\mathrm{AUC}_{k,\mathrm{norm}}=\frac{\mathrm{AUC}_k-\mathrm{AUC}_{k,\mathrm{STL}}^{\mathrm{q10}}}
{\mathrm{AUC}_{k,\mathrm{STL}}^{\mathrm{best}}-\mathrm{AUC}_{k,\mathrm{STL}}^{\mathrm{q10}}}.
\label{eq:normalized-auc}
\end{equation}
Here, $\mathrm{AUC}_k$ is the task's validation AUC, using the same metric as the corresponding STL task: ShotAUC for DP and WindowAUC for ELM, MARFE, and H/L.
The reference values $\mathrm{AUC}_{k,\mathrm{STL}}^{\mathrm{q10}}$ and $\mathrm{AUC}_{k,\mathrm{STL}}^{\mathrm{best}}$ are the 10th percentile and maximum, respectively, of the validation AUCs obtained across all trials for the corresponding STL task.
These fixed reference values map to 0 and 1, respectively, expressing performance relative to each task's own STL performance range.
Using the 10th percentile rather than the minimum reduces sensitivity to exceptionally poor STL results.
Before calculating the geometric mean, each normalized AUC is clipped to $[0,1]$, preventing a gain beyond one task's baseline from compensating for another task's deficit within the first term.
Consequently, the first term equals 1 only when all four tasks match or exceed their corresponding STL baselines.

The second term rewards performance improvements beyond the STL baselines.
When a task's normalized AUC exceeds 1, $\max(0,\mathrm{AUC}_{k,\mathrm{norm}}-1)$ retains the excess; otherwise, that task contributes zero to this term.
Summing these gains across the four tasks with a weight of 0.0001 allows the composite score to reflect further performance improvements even after the first term reaches its upper limit.

The third term provides a small reward for faster models, contributing at most 0.0001 to the total score.
Inference latency is converted into an inference-time score.
The inference latency $t_{\mathrm{inf}}$ is the time required for the model to generate all four task outputs from one input window.
Before the composite score is first calculated in each MTL trial, the model's inference latency is assessed once under uniform hardware, input, and timing conditions using repeated measurements after warm-up.
The resulting $t_{\mathrm{inf}}$ defines the inference-time score:
\begin{equation}
\mathrm{InfTimeScore}=\operatorname{clip}\!\left(
\frac{500\,\mu\mathrm{s}-t_{\mathrm{inf}}}{450\,\mu\mathrm{s}},0,1\right).
\label{eq:speed-score}
\end{equation}
$\mathrm{InfTimeScore}=1$ for inference latencies of $50\,\mu\mathrm{s}$ or less and $\mathrm{InfTimeScore}=0$ for latencies of $500\,\mu\mathrm{s}$ or more, with linear interpolation between these bounds.
These latency bounds are chosen considering the control periods and latency margins of some PCS control algorithms.
The resulting $\mathrm{InfTimeScore}$ remains fixed within the trial and is reused in every first-stage epoch and in the model's score after both training stages, without repeating the timing assessment each epoch.
Because the inference-time score is constant within a trial, it does not change the ranking of that model's checkpoints across epochs, but allows inference time to contribute to comparisons across models.

The composite score guides selection during training and hyperparameter optimization, but a total score above 1 does not itself guarantee that every task reaches its STL baseline; the final assessment of negative transfer therefore uses the individual test AUCs.
At the end of each epoch in the first MTL training stage, $\mathrm{ValScore}_{\mathrm{MTL}}$ is used for checkpoint selection and to determine whether early stopping is required.
If this score does not improve for ten consecutive epochs, the first stage stops and the checkpoint with the highest composite validation score is restored.

The second stage starts from the best first-stage checkpoint, freezes the shared feature extractors, and separately refines each task-specific branch using its own task-label loss $\mathcal{L}_k$ (Figure~\ref{fig:model-architecture}c).
At the end of each epoch, each branch independently selects its checkpoint and checks for early stopping using its own validation ShotAUC for DP or WindowAUC for ELM, MARFE, and H/L.
If a branch's validation metric does not improve for ten consecutive epochs, that branch stops training and restores its best checkpoint; stopped branches remain fixed while the others continue training.
Model training ends when all four branches have stopped.
In the trained model, each task-specific branch uses the parameters from its own best second-stage checkpoint, while the shared feature extractors retain the parameters from the best first-stage checkpoint.

After each trial, the model's validation score guides subsequent hyperparameter proposals, determines when hyperparameter optimization terminates, and provides the criterion for final model selection.
Hyperparameter optimization is conducted independently for each STL task to maximize validation ShotAUC for DP or WindowAUC for ELM, MARFE, and H/L; a separate optimization for the traditional MTL method maximizes $\mathrm{ValScore}_{\mathrm{MTL}}$.
These optimization processes use Optuna~\cite{akiba2019optuna}: hyperparameters are sampled randomly for the first 50 trials and proposed by TPE thereafter.
Each optimization terminates when 100 consecutive trials fail to improve its best validation score or when 400 trials have been completed.
The model with the highest validation score from each optimization is selected, yielding four STL baseline models and one traditional MTL baseline model.

\subsection{Teacher-guided multi-task learning}
\label{subsec:multi-task-ensemble-distillation}
\begin{figure}[!t]
\centering
\captionsetup{font=figonefont,skip=5pt}
\includegraphics[width=\linewidth]{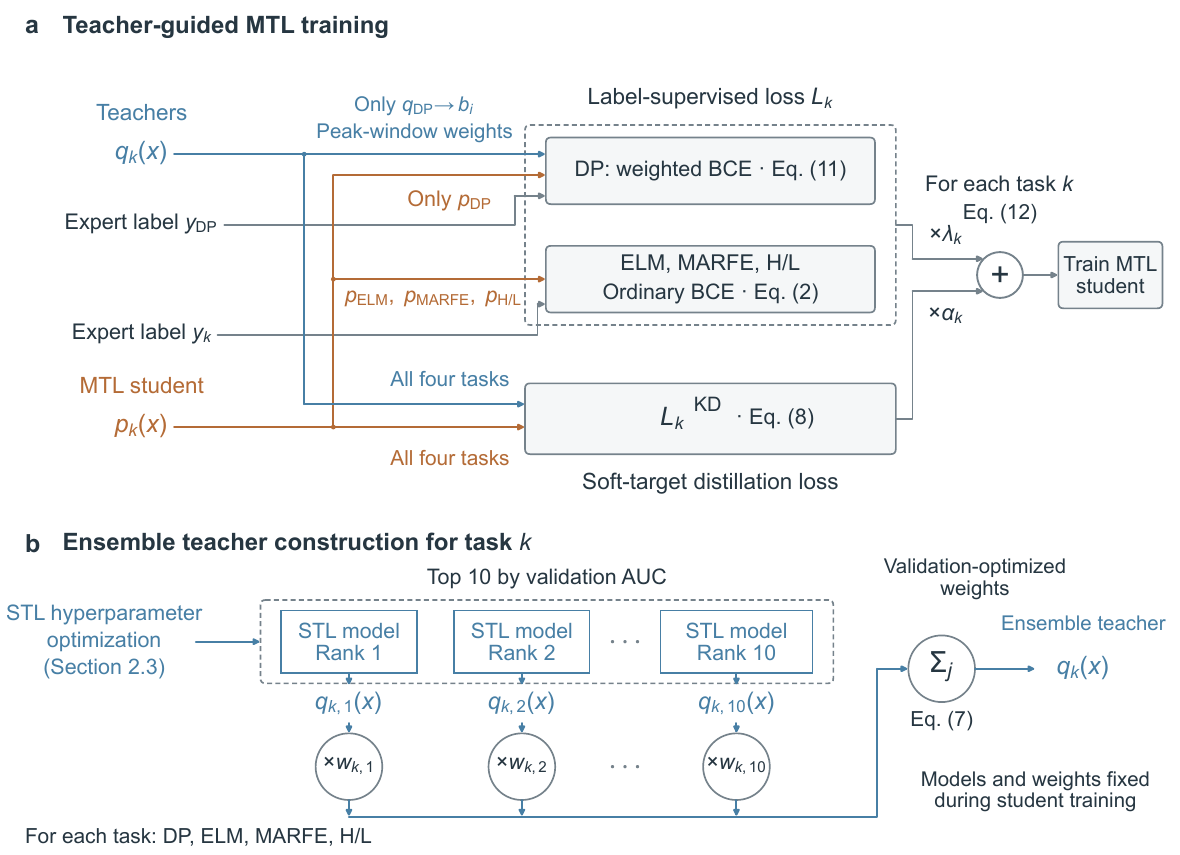}
\caption{Teacher-guided multi-task learning.
(a) Student training with label supervision and soft-target distillation.
(b) Task-specific ensemble teacher construction.
Blue and orange identify the teacher and student sides, respectively.
Crossing paths without a dot are not connected.
}
\label{fig:ensemble-distillation}
\end{figure}

The teacher-guided MTL method modifies the traditional MTL training objective through two forms of teacher guidance (Figure~\ref{fig:ensemble-distillation}).
Soft-target distillation~\cite{hinton2015distilling} adds a loss for each of the four tasks to mitigate negative transfer.
DP peak-disruptivity window weighting modifies the existing DP label-supervised loss by increasing the contributions of teacher-selected windows to further improve DP performance.

Both mechanisms use predictions from task-specific ensemble teachers.
For each task, the STL models trained during hyperparameter optimization (Section~\ref{subsec:baseline-construction}) are ranked by validation AUC, and the top 10 are combined to form an ensemble teacher.
Their combination weights are optimized to maximize validation ShotAUC for DP or WindowAUC for ELM, MARFE, and H/L.
Let $q_{k,j}(x)$ denote the probability predicted by member $j$ for task $k$ and input window $x$, and let $w_{k,j}$ denote its combination weight.
The ensemble teacher predicts the probability $q_k(x)$ for task $k$ as
\begin{equation}
q_k(x)=\sum_{j=1}^{10}w_{k,j}q_{k,j}(x).
\label{eq:ensemble-teacher}
\end{equation}
The ensemble teacher parameters and combination weights remain fixed throughout student training.

\suppressfloats[t]

Soft-target distillation uses each task's teacher probabilities as soft targets for all training windows, including those without a valid expert label for that task.
For task $k$ and input window $x$, $p_k(x)$ denotes the probability predicted by the MTL student model.
Teacher and student probabilities are adjusted by a temperature $\tau$ and denoted by $q_k^{(\tau)}$ and $p_k^{(\tau)}$, respectively.
A binary gate $g_{k,i}$ controls whether window $i$ contributes to distillation.
The distillation loss for task $k$ is
\begin{equation}
\mathcal{L}_k^{\mathrm{KD}}
=\operatorname{mean}_{i}\!\left[
g_{k,i}\tau^2\operatorname{BCE}\!\left(q_k^{(\tau)}(x_{k,i}),p_k^{(\tau)}(x_{k,i})\right)
\right],
\label{eq:distillation-window-loss}
\end{equation}
The mean includes all input windows in the parameter update, including unlabelled windows and windows with $g_{k,i}=0$.

The temperature-adjusted teacher probability is
\begin{equation}
q_k^{(\tau)}(x_{k,i})
=\sigma\!\left(\frac{\operatorname{logit}(q_k(x_{k,i}))}{\tau}\right),
\label{eq:distillation-temperature}
\end{equation}
where $\sigma$ is the Sigmoid function and $\operatorname{logit}(q_k)=\ln[q_k/(1-q_k)]$ is its inverse.
Applying the same transformation to the student probability $p_k(x_{k,i})$ gives $p_k^{(\tau)}(x_{k,i})$.
The temperature is optimized as a hyperparameter and remains fixed during student training in each trial.

Following teacher-error-based distillation formulations~\cite{chen2017efficient,saputra2019distilling}, the gate disables distillation for a window when the student's prediction is closer to a valid expert label than the teacher's; otherwise, distillation remains active.
\par
\begin{samepage}
The gating variable is therefore defined as
\begin{equation}
g_{k,i}=\begin{cases}
0, & \begin{aligned}[t]
&\text{if }y_{k,i}\text{ is valid and}\\
&|p_k(x_{k,i})-y_{k,i}|<|q_k(x_{k,i})-y_{k,i}|,
\end{aligned}\\
1, & \text{otherwise}.
\end{cases}
\label{eq:distillation-gate}
\end{equation}
\end{samepage}
The gate uses probabilities without temperature scaling and does not affect label supervision.

The label-supervised loss $\mathcal{L}_k$ uses only windows with valid expert labels for task $k$.
The label-supervised losses for ELM, MARFE, and H/L retain the ordinary BCE form in Equation~\eqref{eq:baseline-task-loss}, while DP uses the weighted BCE form in Equation~\eqref{eq:dp-key-window-loss}.
For each training shot, the fixed DP ensemble teacher selects the eligible DP training window with the highest predicted disruption probability, termed the peak-disruptivity window.
For disruptive shots, eligible windows must end at least 10\,ms before disruption; all available DP training windows in non-disruptive shots are eligible.
Window selection is completed before student training and remains fixed throughout training.
Teacher-selected peak-disruptivity windows receive label-loss weight $b_i=\rho>1$, while other windows use $b_i=1$, giving
\begin{equation}
\mathcal{L}_{\mathrm{DP}}
=\frac{\sum_{i=1}^{N_{\mathrm{DP}}}b_i\operatorname{BCE}\!\left(y_{\mathrm{DP},i},p_{\mathrm{DP}}(x_{\mathrm{DP},i})\right)}
{\sum_{i=1}^{N_{\mathrm{DP}}}b_i}.
\label{eq:dp-key-window-loss}
\end{equation}
Here, $N_{\mathrm{DP}}$ counts only windows with valid DP labels.

This weighting changes only each window's contribution to the DP label-supervised loss, preserves the original labels, and applies in both training stages.

For each task, the label-supervised loss $\mathcal{L}_k$ and distillation loss $\mathcal{L}_k^{\mathrm{KD}}$ are weighted by $\lambda_k$ and $\alpha_k$, respectively, giving the task training objective
\begin{equation}
\lambda_k\mathcal{L}_k+\alpha_k\mathcal{L}_k^{\mathrm{KD}}.
\label{eq:teacher-guided-task-objective}
\end{equation}

Both MTL methods use the same modular architecture framework and two-stage training procedure.
In the first stage, the four task objectives are summed to train the entire student network.
The second stage freezes the shared feature extractors and refines the task-specific branches separately using their corresponding training objectives.
The validation metrics, early-stopping rules, and checkpoint selection in each stage follow Section~\ref{subsec:baseline-construction}.
After training, the MTL student model generates all four task outputs without running the teachers.

Hyperparameter optimization for the teacher-guided MTL method is conducted independently.
The optimized hyperparameters include those in Section~\ref{subsec:baseline-construction}, together with the distillation-loss weights $\alpha_k$, temperature $\tau$, and DP window multiplier $\rho$.
Their applicability is listed in Table~\ref{tab:baseline-configuration} in Appendix~\ref{app:baseline-configuration}.
The sampling and stopping rules for hyperparameter optimization also follow Section~\ref{subsec:baseline-construction}.
After both training stages, the model with the highest MTL composite validation score $\mathrm{ValScore}_{\mathrm{MTL}}$ across trials is selected as the final teacher-guided MTL model.

\clearpage

\section{Results and analysis}
\label{sec:results}

\subsection{STL baseline models and ensemble teachers}
\label{subsec:results-hpo-validation}
\begin{figure}[!htbp]
\centering
\captionsetup{font=figonefont}
\includegraphics[width=\linewidth]{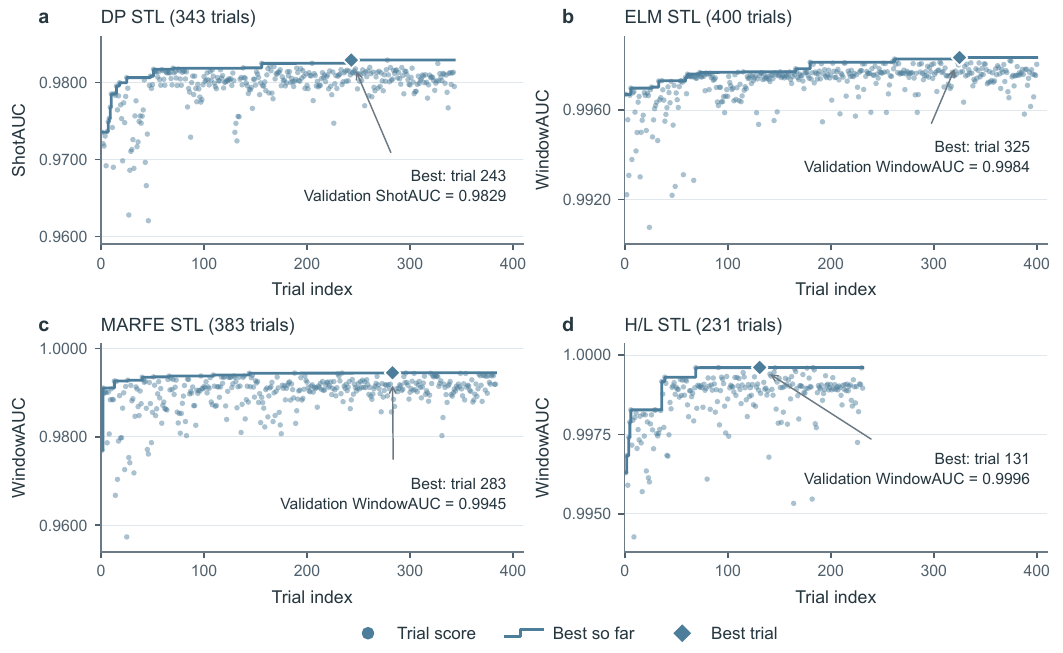}
\caption{STL hyperparameter optimization on validation data: (a) DP; (b) ELM; (c) MARFE; (d) H/L.
Points show trial scores, step lines the best score so far, and diamonds the best trials.
}
\label{fig:hpo-validation}
\end{figure}

For all four tasks, the best validation AUC tended to converge as STL hyperparameter optimization progressed (Figure~\ref{fig:hpo-validation}).
The best validation AUCs for DP, MARFE, and H/L last improved at trials 243, 283, and 131, respectively.
Hyperparameter optimization for these tasks terminated after 100 consecutive trials without improvement, after 343, 383, and 231 trials, respectively.
ELM hyperparameter optimization reached the preset limit of 400 trials.
Its best validation AUC last improved at trial 325, with no further improvement in the remaining 75 trials.
For each task, the model with the highest validation AUC was selected as the STL baseline model.

Table~\ref{tab:stl-baselines} reports independent test AUCs for the four STL baseline models and their corresponding ensemble teachers, assessing validation-selected models on data not used for hyperparameter optimization.
The four STL baseline models achieved test AUCs of 0.9708, 0.9911, 0.9907, and 0.9909, respectively, providing single-task performance references for the subsequent assessment of negative transfer in multi-task learning.

On the same test data, the ensembles achieved higher AUCs than the corresponding STL baseline models for all four tasks (Table~\ref{tab:stl-baselines}), making them better suited to serve as task-specific teachers for MTL student training.
\begingroup
\small
\captionsetup{font=normalsize}
\renewcommand{\arraystretch}{1.18}\setlength{\tabcolsep}{5pt}
\setlength{\LTcapwidth}{\linewidth}
\setcounter{LTchunksize}{200}
\setlength{\LTpre}{4pt}\setlength{\LTpost}{4pt}
\begin{longtable}{@{}>{\raggedright\arraybackslash}p{3.9cm}*{4}{>{\centering\arraybackslash}p{\dimexpr(\linewidth-3.9cm-8\tabcolsep)/4\relax}}@{}}
\caption{Test AUCs of the task-specific STL baseline models and ensemble teachers.}
\label{tab:stl-baselines}
\label{tab:ensemble-teachers}\\
\toprule
Model type & \shortstack{DP\\ShotAUC} & \shortstack{ELM\\WindowAUC} & \shortstack{MARFE\\WindowAUC} & \shortstack{H/L\\WindowAUC} \\
\midrule
\endfirsthead
\multicolumn{5}{@{}l@{}}{\tablename~\thetable\ (continued)}\\[3pt]
\toprule
Model type & \shortstack{DP\\ShotAUC} & \shortstack{ELM\\WindowAUC} & \shortstack{MARFE\\WindowAUC} & \shortstack{H/L\\WindowAUC} \\
\midrule
\endhead
\midrule
\multicolumn{5}{r}{\footnotesize Continued on next page}\\
\endfoot
\bottomrule
\noalign{\vskip4pt}
\multicolumn{5}{@{}l@{}}{\begin{minipage}{\linewidth}\footnotesize\raggedright
Each task has its own STL baseline model and ensemble teacher.
\end{minipage}}\\
\noalign{\vskip0pt}
\endlastfoot
STL baseline models & 0.9708 & 0.9911 & 0.9907 & 0.9909 \\
Ensemble teachers & 0.9719 & 0.9948 & 0.9945 & 0.9967 \\
\end{longtable}
\endgroup

\FloatBarrier

\subsection{Mitigation of negative transfer}
\label{subsec:results-negative-transfer}
\label{subsec:results-distillation}
\begin{figure}[!htbp]
\centering
\captionsetup{font=figonefont}
\includegraphics[width=\linewidth]{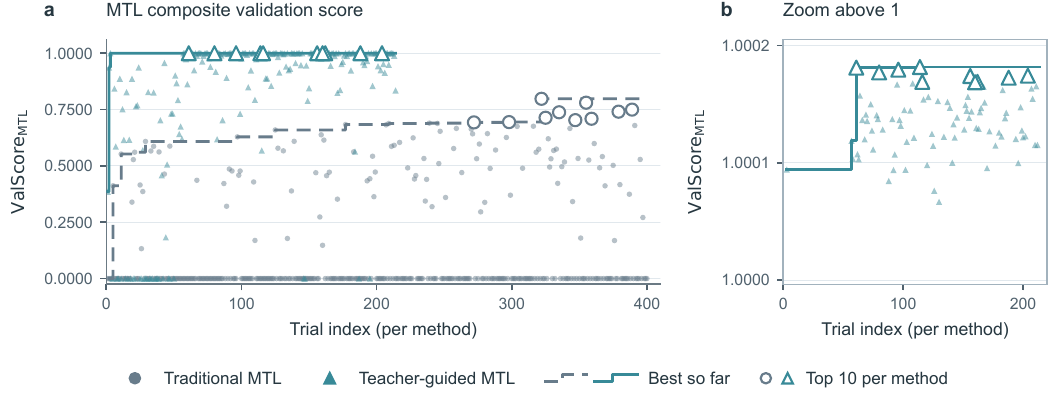}
\caption{MTL hyperparameter optimization on validation data.
(a) Trial scores from hyperparameter optimization for the traditional MTL and teacher-guided MTL methods (Eq.~\eqref{eq:performance-efficiency}).
(b) Zoom of teacher-guided MTL scores between 1 and 1.0002.
Markers show trial scores; step lines show the best score so far for each method.
Outlined markers identify the trials that produced the top 10 models by validation score for each method.
}
\label{fig:mtl-hpo-validation}
\end{figure}

During hyperparameter optimization, the teacher-guided MTL method achieved a higher best MTL composite validation score, $\mathrm{ValScore}_{\mathrm{MTL}}$, than the traditional MTL method (Figure~\ref{fig:mtl-hpo-validation}).
Hyperparameter optimization for the teacher-guided MTL method terminated after 100 consecutive trials without improvement in the best validation score, whereas optimization for the traditional MTL method terminated upon reaching the budget of 400 trials.

For each MTL method, the top 10 models ranked by $\mathrm{ValScore}_{\mathrm{MTL}}$ during hyperparameter optimization were compared in terms of per-task test performance (Table~\ref{tab:mtl-top10-validation-test}).
The results showed that the teacher-guided MTL models eliminated the negative transfer observed in the traditional MTL models.
Specifically, negative transfer was observed in all top 10 traditional MTL models: their DP AUCs were all below the DP STL baseline AUC, whereas their H/L AUCs were all above the H/L STL baseline AUC; ELM and MARFE showed both improvements and declines.
In contrast, all top 10 teacher-guided MTL models exceeded the corresponding STL baseline models in test AUC on all four tasks, showing that their elimination of negative transfer was not a coincidence arising from a particular hyperparameter setting.
In addition, some teacher-guided MTL models even exceeded the corresponding ensemble teachers in test AUC on DP, MARFE, and H/L, whereas all top 10 teacher-guided MTL models had lower ELM test AUCs than the ELM ensemble teacher.
\begingroup
\captionsetup{font=normalsize}

\fontsize{10}{12.8}\selectfont
\renewcommand{\arraystretch}{1.14}\setlength{\tabcolsep}{0pt}
\newcommand{\TableFourAUC}[1]{\makebox[40pt][l]{#1}}
\setlength{\LTcapwidth}{\linewidth}
\setcounter{LTchunksize}{200}
\setlength{\LTpre}{4pt}\setlength{\LTpost}{4pt}
\begin{longtable}{@{}>{\centering\arraybackslash}p{105pt}@{\hspace{5pt}}*{4}{>{\centering\arraybackslash}p{60pt}@{\hspace{5pt}}}>{\centering\arraybackslash}p{\dimexpr\linewidth-370pt\relax}@{}}
\caption{Test AUCs and negative-transfer status of the top 10 models by validation score for each MTL method.}
\label{tab:mtl-top10-validation-test}
\label{tab:mtl-top10-test-wt}\\
\toprule
\shortstack{Rank by\\$\mathrm{ValScore}_{\mathrm{MTL}}$} & \shortstack{DP\\ShotAUC} & \shortstack{ELM\\WindowAUC} & \shortstack{MARFE\\WindowAUC} & \shortstack{H/L\\WindowAUC} & \shortstack{Negative\\transfer} \\
\midrule
\endfirsthead
\multicolumn{6}{@{}l@{}}{\tablename~\thetable\ (continued)}\\[3pt]
\toprule
\shortstack{Rank by\\$\mathrm{ValScore}_{\mathrm{MTL}}$} & \shortstack{DP\\ShotAUC} & \shortstack{ELM\\WindowAUC} & \shortstack{MARFE\\WindowAUC} & \shortstack{H/L\\WindowAUC} & \shortstack{Negative\\transfer} \\
\midrule
\endhead
\midrule
\multicolumn{6}{r}{\footnotesize Continued on next page}\\
\endfoot
\bottomrule
\noalign{\vskip4pt}
\multicolumn{6}{@{}l@{}}{\begin{minipage}{\linewidth}\fontsize{9}{10.8}\selectfont
Models are ranked by $\mathrm{ValScore}_{\mathrm{MTL}}$ (Eq.~\eqref{eq:performance-efficiency}).
Negative transfer is recorded as Observed if any task AUC is below that of its corresponding STL baseline model; otherwise, it is recorded as Not observed.
Blue $\uparrow$: above STL but not above the ensemble teacher; blue $\uparrow\uparrow$: above the teacher; orange $\downarrow$: below STL.
Comparisons use unrounded AUCs.
\end{minipage}}\\
\noalign{\vskip0pt}
\endlastfoot
\multicolumn{6}{l}{\textbf{Traditional MTL: top 10 models}}\\*\addlinespace[2pt]
1 & \TableFourAUC{\textcolor[HTML]{D55E00}{0.9642\,\ensuremath{\downarrow}}} & \TableFourAUC{\textcolor[HTML]{0072B2}{0.9953\,\ensuremath{\uparrow\uparrow}}} & \TableFourAUC{\textcolor[HTML]{D55E00}{0.9904\,\ensuremath{\downarrow}}} & \TableFourAUC{\textcolor[HTML]{0072B2}{0.9987\,\ensuremath{\uparrow\uparrow}}} & Observed \\*
2 & \TableFourAUC{\textcolor[HTML]{D55E00}{0.9590\,\ensuremath{\downarrow}}} & \TableFourAUC{\textcolor[HTML]{0072B2}{0.9936\,\ensuremath{\uparrow}}} & \TableFourAUC{\textcolor[HTML]{0072B2}{0.9937\,\ensuremath{\uparrow}}} & \TableFourAUC{\textcolor[HTML]{0072B2}{0.9989\,\ensuremath{\uparrow\uparrow}}} & Observed \\*
3 & \TableFourAUC{\textcolor[HTML]{D55E00}{0.9644\,\ensuremath{\downarrow}}} & \TableFourAUC{\textcolor[HTML]{0072B2}{0.9930\,\ensuremath{\uparrow}}} & \TableFourAUC{\textcolor[HTML]{0072B2}{0.9918\,\ensuremath{\uparrow}}} & \TableFourAUC{\textcolor[HTML]{0072B2}{0.9989\,\ensuremath{\uparrow\uparrow}}} & Observed \\*
4 & \TableFourAUC{\textcolor[HTML]{D55E00}{0.9590\,\ensuremath{\downarrow}}} & \TableFourAUC{\textcolor[HTML]{0072B2}{0.9917\,\ensuremath{\uparrow}}} & \TableFourAUC{\textcolor[HTML]{0072B2}{0.9932\,\ensuremath{\uparrow}}} & \TableFourAUC{\textcolor[HTML]{0072B2}{0.9988\,\ensuremath{\uparrow\uparrow}}} & Observed \\*
5 & \TableFourAUC{\textcolor[HTML]{D55E00}{0.9644\,\ensuremath{\downarrow}}} & \TableFourAUC{\textcolor[HTML]{0072B2}{0.9920\,\ensuremath{\uparrow}}} & \TableFourAUC{\textcolor[HTML]{0072B2}{0.9930\,\ensuremath{\uparrow}}} & \TableFourAUC{\textcolor[HTML]{0072B2}{0.9989\,\ensuremath{\uparrow\uparrow}}} & Observed \\*
6 & \TableFourAUC{\textcolor[HTML]{D55E00}{0.9598\,\ensuremath{\downarrow}}} & \TableFourAUC{\textcolor[HTML]{D55E00}{0.9910\,\ensuremath{\downarrow}}} & \TableFourAUC{\textcolor[HTML]{D55E00}{0.9893\,\ensuremath{\downarrow}}} & \TableFourAUC{\textcolor[HTML]{0072B2}{0.9986\,\ensuremath{\uparrow\uparrow}}} & Observed \\*
7 & \TableFourAUC{\textcolor[HTML]{D55E00}{0.9630\,\ensuremath{\downarrow}}} & \TableFourAUC{\textcolor[HTML]{0072B2}{0.9920\,\ensuremath{\uparrow}}} & \TableFourAUC{\textcolor[HTML]{D55E00}{0.9884\,\ensuremath{\downarrow}}} & \TableFourAUC{\textcolor[HTML]{0072B2}{0.9985\,\ensuremath{\uparrow\uparrow}}} & Observed \\*
8 & \TableFourAUC{\textcolor[HTML]{D55E00}{0.9549\,\ensuremath{\downarrow}}} & \TableFourAUC{\textcolor[HTML]{0072B2}{0.9919\,\ensuremath{\uparrow}}} & \TableFourAUC{\textcolor[HTML]{D55E00}{0.9897\,\ensuremath{\downarrow}}} & \TableFourAUC{\textcolor[HTML]{0072B2}{0.9986\,\ensuremath{\uparrow\uparrow}}} & Observed \\*
9 & \TableFourAUC{\textcolor[HTML]{D55E00}{0.9613\,\ensuremath{\downarrow}}} & \TableFourAUC{\textcolor[HTML]{0072B2}{0.9924\,\ensuremath{\uparrow}}} & \TableFourAUC{\textcolor[HTML]{0072B2}{0.9942\,\ensuremath{\uparrow}}} & \TableFourAUC{\textcolor[HTML]{0072B2}{0.9986\,\ensuremath{\uparrow\uparrow}}} & Observed \\*
10 & \TableFourAUC{\textcolor[HTML]{D55E00}{0.9622\,\ensuremath{\downarrow}}} & \TableFourAUC{\textcolor[HTML]{0072B2}{0.9940\,\ensuremath{\uparrow}}} & \TableFourAUC{\textcolor[HTML]{0072B2}{0.9921\,\ensuremath{\uparrow}}} & \TableFourAUC{\textcolor[HTML]{0072B2}{0.9993\,\ensuremath{\uparrow\uparrow}}} & Observed \\
\noalign{\penalty0}
\midrule
\multicolumn{6}{l}{\textbf{Teacher-guided MTL: top 10 models}}\\*\addlinespace[2pt]
1 & \TableFourAUC{\textcolor[HTML]{0072B2}{0.9734\,\ensuremath{\uparrow\uparrow}}} & \TableFourAUC{\textcolor[HTML]{0072B2}{0.9928\,\ensuremath{\uparrow}}} & \TableFourAUC{\textcolor[HTML]{0072B2}{0.9952\,\ensuremath{\uparrow\uparrow}}} & \TableFourAUC{\textcolor[HTML]{0072B2}{0.9981\,\ensuremath{\uparrow\uparrow}}} & Not observed \\*
2 & \TableFourAUC{\textcolor[HTML]{0072B2}{0.9722\,\ensuremath{\uparrow\uparrow}}} & \TableFourAUC{\textcolor[HTML]{0072B2}{0.9934\,\ensuremath{\uparrow}}} & \TableFourAUC{\textcolor[HTML]{0072B2}{0.9948\,\ensuremath{\uparrow\uparrow}}} & \TableFourAUC{\textcolor[HTML]{0072B2}{0.9977\,\ensuremath{\uparrow\uparrow}}} & Not observed \\*
3 & \TableFourAUC{\textcolor[HTML]{0072B2}{0.9741\,\ensuremath{\uparrow\uparrow}}} & \TableFourAUC{\textcolor[HTML]{0072B2}{0.9934\,\ensuremath{\uparrow}}} & \TableFourAUC{\textcolor[HTML]{0072B2}{0.9944\,\ensuremath{\uparrow}}} & \TableFourAUC{\textcolor[HTML]{0072B2}{0.9979\,\ensuremath{\uparrow\uparrow}}} & Not observed \\*
4 & \TableFourAUC{\textcolor[HTML]{0072B2}{0.9738\,\ensuremath{\uparrow\uparrow}}} & \TableFourAUC{\textcolor[HTML]{0072B2}{0.9930\,\ensuremath{\uparrow}}} & \TableFourAUC{\textcolor[HTML]{0072B2}{0.9933\,\ensuremath{\uparrow}}} & \TableFourAUC{\textcolor[HTML]{0072B2}{0.9978\,\ensuremath{\uparrow\uparrow}}} & Not observed \\*
5 & \TableFourAUC{\textcolor[HTML]{0072B2}{0.9715\,\ensuremath{\uparrow}}} & \TableFourAUC{\textcolor[HTML]{0072B2}{0.9941\,\ensuremath{\uparrow}}} & \TableFourAUC{\textcolor[HTML]{0072B2}{0.9949\,\ensuremath{\uparrow\uparrow}}} & \TableFourAUC{\textcolor[HTML]{0072B2}{0.9979\,\ensuremath{\uparrow\uparrow}}} & Not observed \\*
6 & \TableFourAUC{\textcolor[HTML]{0072B2}{0.9720\,\ensuremath{\uparrow\uparrow}}} & \TableFourAUC{\textcolor[HTML]{0072B2}{0.9936\,\ensuremath{\uparrow}}} & \TableFourAUC{\textcolor[HTML]{0072B2}{0.9937\,\ensuremath{\uparrow}}} & \TableFourAUC{\textcolor[HTML]{0072B2}{0.9982\,\ensuremath{\uparrow\uparrow}}} & Not observed \\*
7 & \TableFourAUC{\textcolor[HTML]{0072B2}{0.9736\,\ensuremath{\uparrow\uparrow}}} & \TableFourAUC{\textcolor[HTML]{0072B2}{0.9945\,\ensuremath{\uparrow}}} & \TableFourAUC{\textcolor[HTML]{0072B2}{0.9946\,\ensuremath{\uparrow\uparrow}}} & \TableFourAUC{\textcolor[HTML]{0072B2}{0.9985\,\ensuremath{\uparrow\uparrow}}} & Not observed \\*
8 & \TableFourAUC{\textcolor[HTML]{0072B2}{0.9735\,\ensuremath{\uparrow\uparrow}}} & \TableFourAUC{\textcolor[HTML]{0072B2}{0.9944\,\ensuremath{\uparrow}}} & \TableFourAUC{\textcolor[HTML]{0072B2}{0.9945\,\ensuremath{\uparrow}}} & \TableFourAUC{\textcolor[HTML]{0072B2}{0.9986\,\ensuremath{\uparrow\uparrow}}} & Not observed \\*
9 & \TableFourAUC{\textcolor[HTML]{0072B2}{0.9718\,\ensuremath{\uparrow}}} & \TableFourAUC{\textcolor[HTML]{0072B2}{0.9944\,\ensuremath{\uparrow}}} & \TableFourAUC{\textcolor[HTML]{0072B2}{0.9947\,\ensuremath{\uparrow\uparrow}}} & \TableFourAUC{\textcolor[HTML]{0072B2}{0.9976\,\ensuremath{\uparrow\uparrow}}} & Not observed \\*
10 & \TableFourAUC{\textcolor[HTML]{0072B2}{0.9723\,\ensuremath{\uparrow\uparrow}}} & \TableFourAUC{\textcolor[HTML]{0072B2}{0.9927\,\ensuremath{\uparrow}}} & \TableFourAUC{\textcolor[HTML]{0072B2}{0.9939\,\ensuremath{\uparrow}}} & \TableFourAUC{\textcolor[HTML]{0072B2}{0.9980\,\ensuremath{\uparrow\uparrow}}} & Not observed \\
\noalign{\penalty0}
\midrule
\multicolumn{1}{@{}l}{\textbf{STL baseline models}} & \TableFourAUC{0.9708} & \TableFourAUC{0.9911} & \TableFourAUC{0.9907} & \TableFourAUC{0.9909} & -- \\
\multicolumn{1}{@{}l}{\textbf{Ensemble teachers}} & \TableFourAUC{0.9719} & \TableFourAUC{0.9948} & \TableFourAUC{0.9945} & \TableFourAUC{0.9967} & -- \\
\end{longtable}
\endgroup

\FloatBarrier

For each STL task and each MTL method, the model ranked first by validation score in its respective hyperparameter optimization was retrained across fifteen seeds with its hyperparameters fixed.
This retraining yielded higher mean test AUCs for the teacher-guided MTL method than for the corresponding STL method on all four tasks (Table~\ref{tab:training-variability}).
Mean test AUCs for the teacher-guided MTL method also exceeded those for the traditional MTL method on DP, ELM, and MARFE.
Per-seed results are reported in Appendix~\ref{app:repeat15}.
\begingroup
\small
\captionsetup{font=normalsize}

\renewcommand{\arraystretch}{1.2}\setlength{\tabcolsep}{2.5pt}
\setlength{\LTcapwidth}{\linewidth}
\setcounter{LTchunksize}{200}
\setlength{\LTpre}{4pt}\setlength{\LTpost}{4pt}
\begin{longtable}{@{}lcccc@{}}
\caption{Test AUCs across fifteen training seeds with validation-selected hyperparameters.}
\label{tab:training-variability}\\
\toprule
& \multicolumn{4}{c}{\textbf{Test AUC (mean $\pm$ sample SD)}} \\
\cmidrule(l){2-5}
\textbf{Method} & \shortstack{DP\\ShotAUC} & \shortstack{ELM\\WindowAUC} & \shortstack{MARFE\\WindowAUC} & \shortstack{H/L\\WindowAUC} \\
\midrule
\endfirsthead
\multicolumn{5}{@{}l@{}}{\tablename~\thetable\ (continued)}\\[3pt]
\toprule
& \multicolumn{4}{c}{\textbf{Test AUC (mean $\pm$ sample SD)}} \\
\cmidrule(l){2-5}
\textbf{Method} & \shortstack{DP\\ShotAUC} & \shortstack{ELM\\WindowAUC} & \shortstack{MARFE\\WindowAUC} & \shortstack{H/L\\WindowAUC} \\
\midrule
\endhead
\midrule
\multicolumn{5}{r}{\footnotesize Continued on next page}\\
\endfoot
\bottomrule
\noalign{\vskip4pt}
\multicolumn{5}{@{}l@{}}{\begin{minipage}{\linewidth}\footnotesize\raggedright
Retraining fixes the hyperparameters of the model ranked first by validation score for each MTL method and of the STL baseline model for each task.
Results are mean $\pm$ sample SD across fifteen training seeds, without repeating hyperparameter optimization.
\end{minipage}}\\
\noalign{\vskip0pt}
\endlastfoot
Teacher-guided MTL & $0.9732 \pm 0.0019$ & $0.9934 \pm 0.0008$ & $0.9942 \pm 0.0006$ & $0.9978 \pm 0.0003$ \\
\midrule
Traditional MTL & $0.9647 \pm 0.0020$ & $0.9913 \pm 0.0016$ & $0.9881 \pm 0.0023$ & $0.9983 \pm 0.0007$ \\
\midrule
STL & $0.9674 \pm 0.0018$ & $0.9926 \pm 0.0013$ & $0.9909 \pm 0.0007$ & $0.9918 \pm 0.0068$ \\
\end{longtable}
\endgroup

\FloatBarrier

In addition, warning time (WT) is not a primary metric of this study and is not used to assess negative transfer. Nevertheless, Appendix~\ref{app:warning-time-distributions} provides a descriptive comparison of WT for the fifteen training repeats reported in Table~\ref{tab:training-variability}.

\subsection{Ablation analysis}
\label{subsec:results-ablation}
\label{subsec:repeat-ablation-protocol}
For the teacher-guided MTL model ranked first by validation score, ablation analysis showed that soft-target distillation improved mean test AUC for all four tasks, while DP peak-disruptivity window weighting further improved mean DP ShotAUC (Table~\ref{tab:component-effects}).
The model architecture was fixed while all four combinations of retaining or removing the two components were compared.
All conditions retained label supervision and used the same fifteen training seeds for paired comparisons.

Adding soft-target distillation improved mean test AUC for all four tasks, both with and without DP window weighting.
For example, with DP window weighting retained, the paired MARFE AUC increase was $0.0104\pm0.0050$ (mean $\pm$ sample SD).

The benefits of DP window weighting were concentrated on DP.
With distillation retained, adding this component increased DP ShotAUC in all fifteen paired training seeds, with a mean increase of $0.0025\pm0.0014$.
The absolute mean AUC changes for the other three tasks were no greater than 0.0003.
Without distillation, DP window weighting still improved mean DP AUC but reduced mean AUC for the other three tasks.
Per-seed test AUCs for all four conditions are reported in Appendix~\ref{app:ablation15} (Table~\ref{tab:ablation15-per-seed}).
\begingroup
\small
\captionsetup{font=normalsize}

\renewcommand{\arraystretch}{1.2}\setlength{\tabcolsep}{3pt}
\setlength{\LTcapwidth}{\linewidth}
\setcounter{LTchunksize}{200}
\setlength{\LTpre}{4pt}\setlength{\LTpost}{4pt}
\begin{longtable}{@{}>{\raggedright\arraybackslash}m{3.9cm}cccc@{}}
\caption{Ablation results across fifteen paired training seeds.}
\label{tab:component-effects}\\
\toprule
& \multicolumn{4}{c}{\textbf{Test AUC (mean $\pm$ sample SD)}} \\
\cmidrule(l){2-5}
\textbf{Training condition} & \shortstack{DP\\ShotAUC} & \shortstack{ELM\\WindowAUC} & \shortstack{MARFE\\WindowAUC} & \shortstack{H/L\\WindowAUC} \\
\midrule
\endfirsthead
\multicolumn{5}{@{}l@{}}{\tablename~\thetable\ (continued)}\\[3pt]
\toprule
& \multicolumn{4}{c}{\textbf{Test AUC (mean $\pm$ sample SD)}} \\
\cmidrule(l){2-5}
\textbf{Training condition} & \shortstack{DP\\ShotAUC} & \shortstack{ELM\\WindowAUC} & \shortstack{MARFE\\WindowAUC} & \shortstack{H/L\\WindowAUC} \\
\midrule
\endhead
\midrule
\multicolumn{5}{r}{\footnotesize Continued on next page}\\
\endfoot
\bottomrule
\noalign{\vskip4pt}
\multicolumn{5}{@{}l@{}}{\begin{minipage}{\linewidth}\footnotesize\raggedright
All conditions retain label supervision and use the fixed architecture of the teacher-guided MTL model.
``Without both teacher-guidance mechanisms'' removes soft-target distillation and DP peak-disruptivity window weighting.
\end{minipage}}\\
\noalign{\vskip0pt}
\endlastfoot
Teacher-guided MTL\newline (full) & $0.9732 \pm 0.0019$ & $0.9934 \pm 0.0008$ & $0.9942 \pm 0.0006$ & $0.9978 \pm 0.0003$ \\
\addlinespace[3pt]
Without DP\newline peak-disruptivity\newline window weighting & $0.9707 \pm 0.0014$ & $0.9931 \pm 0.0008$ & $0.9942 \pm 0.0006$ & $0.9978 \pm 0.0003$ \\
\addlinespace[3pt]
Without soft-target\newline distillation & $0.9708 \pm 0.0033$ & $0.9912 \pm 0.0023$ & $0.9838 \pm 0.0049$ & $0.9966 \pm 0.0008$ \\
\addlinespace[3pt]
Without both\newline teacher-guidance\newline mechanisms & $0.9587 \pm 0.0043$ & $0.9919 \pm 0.0027$ & $0.9866 \pm 0.0034$ & $0.9968 \pm 0.0010$ \\
\end{longtable}
\endgroup

\FloatBarrier

\subsection{Inference time}
\label{subsec:results-efficiency}
\label{subsec:inference-timing}
\begingroup\emergencystretch=2em
On the same single GPU or single physical CPU core, the top 10 teacher-guided MTL models by validation score required less inference time to generate all four task outputs than the four STL baseline models, while also using fewer parameters (Table~\ref{tab:efficiency-main}).
Measurements used FP32 precision and a batch size of~1.
GPU measurements used a single RTX~4090; CPU measurements used one physical core of an Intel Xeon Gold~6544Y with ONNX Runtime~1.23.2.
GPU inference time was used to calculate the inference-time score during MTL hyperparameter optimization.

Under these conditions, the teacher-guided MTL model ranked first by $\mathrm{ValScore}_{\mathrm{MTL}}$ achieved $2.3$-fold GPU and $1.9$-fold CPU speedups over the four STL baseline models (Table~\ref{tab:efficiency-main}), with fewer than one quarter of their combined parameters.
Specifically, median inference time decreased from $0.513$ to $0.219\,\mathrm{ms}$ on GPU and from $8.124$ to $4.182\,\mathrm{ms}$ on CPU.

Among the top 10 teacher-guided MTL models ranked by $\mathrm{ValScore}_{\mathrm{MTL}}$, the model ranked tenth had the shortest inference time on both GPU and CPU (Table~\ref{tab:efficiency-main}).
This model achieved $5.0$-fold GPU and $8.6$-fold CPU speedups over the four STL baseline models while retaining higher test AUCs on all four tasks (Table~\ref{tab:mtl-top10-validation-test}).

On both GPU and CPU, the ensemble teachers required substantially longer inference times than the corresponding STL baseline models for all four tasks (Table~\ref{tab:efficiency-main}), limiting their suitability for real-time plasma monitoring.
In this study, the ensemble teachers are therefore used only to train the MTL student model, which alone is used for inference.
\par\endgroup

\begingroup
\fontsize{9.5}{11.4}\selectfont
\setlength{\tabcolsep}{0pt}
\renewcommand{\arraystretch}{1.02}
\setlength{\LTcapwidth}{\linewidth}
\setcounter{LTchunksize}{200}
\setlength{\LTpre}{4pt}\setlength{\LTpost}{4pt}
\begin{longtable}{@{}>{\raggedright\arraybackslash}p{\dimexpr\linewidth-349pt\relax}@{\hspace{3pt}}>{\raggedleft\arraybackslash}p{35pt}@{\hspace{3pt}}>{\raggedleft\arraybackslash}p{31pt}@{\hspace{6pt}}>{\fontsize{9.5}{12}\selectfont\raggedleft\arraybackslash}p{61pt}@{\hspace{3pt}}>{\raggedleft\arraybackslash}p{36pt}@{\hspace{3pt}}>{\raggedleft\arraybackslash}p{36pt}@{\hspace{6pt}}>{\fontsize{9.5}{12}\selectfont\raggedleft\arraybackslash}p{73pt}@{\hspace{3pt}}>{\raggedleft\arraybackslash}p{50pt}@{}}
\caption{GPU and CPU inference times and parameter counts.}
\label{tab:efficiency-main}\label{tab:efficiency-main-cpu-onnx}\\
\toprule
\shortstack[l]{Model or\\deployment} & \multicolumn{3}{c}{Single GPU (ms)} & \multicolumn{3}{c}{Single physical CPU core (ms)} & \shortstack{Parameters\\(million)} \\
\cmidrule(lr){2-4}\cmidrule(lr){5-7}
 & Median & P95 & Range & Median & P95 & Range & \\
\midrule
\endfirsthead
\multicolumn{8}{@{}l@{}}{\tablename~\thetable\ (continued)}\\[3pt]
\toprule
\shortstack[l]{Model or\\deployment} & \multicolumn{3}{c}{Single GPU (ms)} & \multicolumn{3}{c}{Single physical CPU core (ms)} & \shortstack{Parameters\\(million)} \\
\cmidrule(lr){2-4}\cmidrule(lr){5-7}
 & Median & P95 & Range & Median & P95 & Range & \\
\midrule
\endhead
\midrule
\multicolumn{8}{r}{\footnotesize Continued on next page}\\
\endfoot
\bottomrule
\noalign{\vskip4pt}
\multicolumn{8}{@{}l@{}}{\begin{minipage}{\linewidth}\fontsize{9.5}{11.4}\selectfont
Median and P95 report the medians of the corresponding estimates from three processes; P95 denotes the 95th percentile of inference time.
Range gives the minimum and maximum of the three process medians.
Bold marks the shortest median inference time among the top 10 teacher-guided MTL models and the corresponding validation rank.
\end{minipage}}\\
\endlastfoot
\multicolumn{8}{@{}l}{\textbf{Teacher-guided MTL: top 10 by $\mathrm{ValScore}_{\mathrm{MTL}}$}} \\*
Rank 1 & 0.219 & 0.225 & 0.213--0.221 & 4.182 & 4.376 & 4.173--4.209 & 0.688 \\
Rank 2 & 0.103 & 0.106 & 0.101--0.103 & 1.342 & 1.371 & 1.335--1.351 & 0.352 \\
Rank 3 & 0.202 & 0.208 & 0.198--0.204 & 2.789 & 3.054 & 2.783--2.791 & 1.011 \\
Rank 4 & 0.239 & 0.245 & 0.235--0.239 & 3.026 & 3.181 & 3.025--3.029 & 0.594 \\
Rank 5 & 0.135 & 0.140 & 0.132--0.137 & 1.027 & 1.082 & 1.025--1.031 & 0.720 \\
Rank 6 & 0.221 & 0.227 & 0.214--0.222 & 4.196 & 4.405 & 4.188--4.198 & 0.669 \\
Rank 7 & 0.202 & 0.208 & 0.195--0.207 & 2.244 & 2.462 & 2.242--2.248 & 1.615 \\
Rank 8 & 0.135 & 0.140 & 0.133--0.135 & 0.972 & 1.031 & 0.970--0.973 & 0.481 \\
Rank 9 & 0.152 & 0.153 & 0.152--0.153 & 2.123 & 2.316 & 2.115--2.123 & 1.642 \\
\textbf{Rank 10} & \textbf{0.102} & 0.106 & 0.102--0.103 & \textbf{0.940} & 0.978 & 0.934--0.944 & 0.401 \\
\midrule
\multicolumn{8}{@{}l}{\textbf{STL baseline models}} \\*
DP & 0.203 & 0.206 & 0.200--0.204 & 2.555 & 2.753 & 2.548--2.556 & 0.847 \\
ELM & 0.223 & 0.229 & 0.219--0.225 & 2.504 & 2.723 & 2.502--2.510 & 1.131 \\
MARFE & 0.158 & 0.163 & 0.153--0.158 & 0.892 & 0.945 & 0.891--0.897 & 0.584 \\
H/L & 0.168 & 0.174 & 0.162--0.170 & 2.047 & 2.106 & 2.046--2.051 & 0.512 \\
\shortstack[l]{Four STL\\baseline models} & 0.513 & 0.535 & 0.512--0.533 & 8.124 & 8.296 & 8.124--8.137 & 3.074 \\
\midrule
\multicolumn{8}{@{}l}{\textbf{Ensemble teachers}} \\*
DP & 1.802 & 1.816 & 1.784--1.821 & 25.327 & 25.579 & 25.297--25.434 & 8.800 \\
ELM & 2.241 & 2.259 & 2.164--2.243 & 26.515 & 26.818 & 26.471--26.754 & 11.335 \\
MARFE & 1.725 & 1.741 & 1.672--1.734 & 22.855 & 23.083 & 22.766--23.027 & 6.130 \\
H/L & 1.679 & 1.696 & 1.630--1.681 & 20.372 & 20.477 & 20.354--20.392 & 5.079 \\
\shortstack[l]{Four teachers\\(40 members)} & 7.882 & 8.013 & 7.859--10.121 & 104.361 & 104.937 & 104.231--104.383 & 31.343 \\
\end{longtable}
\endgroup

\FloatBarrier

\par

\section{Summary}
\label{sec:conclusions}\label{sec:limitations}
This study shows that the teacher-guided MTL method outperforms the corresponding STL method in test-set ROC-AUC for all four tasks, eliminating the negative transfer observed with the traditional MTL method.
On the same single GPU or single physical CPU core, the MTL student model also reduces the inference time required to generate all four task outputs compared with the four STL baseline models.
Ablation results show that soft-target distillation improves mean test AUC on all four tasks, while DP peak-disruptivity window weighting further improves DP performance.
These findings show that the teacher-guided MTL method can outperform the corresponding STL method on every task, even when the same monitoring task combination exhibits negative transfer under the traditional MTL method.
This offers a promising route to plasma monitoring systems that cover a broader range of phenomena, improve per-task performance and reduce inference time with limited real-time computational resources, supporting future high-performance plasma operation.
The current evidence is limited to these four tasks on EAST.
Future work should evaluate the method on more tasks and other devices to assess its broader applicability.

\section*{Acknowledgment}
This work is supported by the National MCF Energy R\&D Program of China (Grant Nos. 2024YFE03020003 and 2022YFE03010002), the National Natural Science Foundation of China (Grant Nos. 12575246, 12575247, U24A20342, 12305255, 12205195, and 12205336), the Institute of Energy, Hefei Comprehensive National Science Center under Grant No. 24KZS304, and funding support from the International Atomic Energy Agency under Research Contract No. 26478.

\clearpage
\appendix
\section{Hyperparameters and optimization scope}
\label{app:baseline-configuration}
Table~\ref{tab:baseline-configuration} summarizes the hyperparameters, their meanings, and the methods in which they are optimized.
\begingroup
\small\fontsize{10.95}{12.2}\selectfont
\setlength{\tabcolsep}{3pt}
\renewcommand{\arraystretch}{1.08}
\setlength{\LTcapwidth}{\linewidth}
\setcounter{LTchunksize}{200}
\setlength{\LTpre}{2pt}\setlength{\LTpost}{2pt}
\begin{longtable}{@{}>{\raggedright\arraybackslash}p{0.25\textwidth}>{\raggedright\arraybackslash}p{\dimexpr0.51\textwidth-4\tabcolsep\relax}>{\raggedright\arraybackslash}p{0.24\textwidth}@{}}
\caption{Hyperparameters, their meanings, and optimization scope.}
\label{tab:baseline-configuration}\\
\toprule
Hyperparameter & Meaning & Optimized in \\
\midrule
\endfirsthead
\multicolumn{3}{@{}l@{}}{\tablename~\thetable\ (continued)}\\[3pt]
\toprule
Hyperparameter & Meaning & Optimized in \\
\midrule
\endhead
\midrule
\multicolumn{3}{r}{\footnotesize Continued on next page}\\
\endfoot
\bottomrule
\noalign{\vskip4pt}
\multicolumn{3}{@{}l@{}}{\begin{minipage}{\textwidth}\footnotesize
The two MTL methods are the traditional MTL method and the teacher-guided MTL method.
Fixed quantities are not optimized.
\end{minipage}}\\
\noalign{\vskip0pt}
\endlastfoot
Signal-group masks~$m_{k,g}$ & Select the input signal groups for task $k$. & ELM, MARFE and H/L STL; both MTL methods; fixed~for~DP. \\
\specialrule{0.25pt}{0pt}{0pt}
Feature-extractor channel counts & Set the output channels of the convolution stages; the first two stages share a channel count. & STL and both MTL methods. \\
\specialrule{0.25pt}{0pt}{0pt}
Convolution kernel sizes & Set the extent of convolution along the signal and time axes, as applicable to each feature extractor. & STL and both MTL methods. \\
\specialrule{0.25pt}{0pt}{0pt}
Projection channels~$C_k$ & Set the common channel count of projected signal-group features for task $k$. & STL and both MTL methods. \\
\specialrule{0.25pt}{0pt}{0pt}
Task-head hidden width~$H_k$ & Set the number of hidden-layer neurons in the task head. & STL and both MTL methods. \\
\specialrule{0.25pt}{0pt}{0pt}
AdamW learning rate & Control the step size of model-parameter updates. & STL and both MTL methods. \\
\specialrule{0.25pt}{0pt}{0pt}
Weight-decay setting & Determine whether weight decay is applied and its strength. & STL and both MTL methods. \\
\specialrule{0.25pt}{0pt}{0pt}
Task-loss weights~$\lambda_k$ & Scale the task-specific label-supervised losses. & Both MTL methods. \\
\specialrule{0.25pt}{0pt}{0pt}
Distillation-loss weights~$\alpha_k$ & Scale the task-specific soft-target distillation losses. & Teacher-guided MTL. \\
\specialrule{0.25pt}{0pt}{0pt}
Distillation temperature~$\tau$ & Control the softness of teacher and student probabilities used in distillation. & Teacher-guided MTL. \\
\specialrule{0.25pt}{0pt}{0pt}
DP window multiplier~$\rho$ & Increase the weight of teacher-selected peak-disruptivity windows in the DP label-supervised loss. & Teacher-guided MTL. \\
\end{longtable}
\endgroup

\FloatBarrier
\begingroup\emergencystretch=2em
\section{Fifteen-seed retraining with fixed hyperparameters}\label{app:repeat15}
\begingroup
\footnotesize

\setlength{\tabcolsep}{2pt}\renewcommand{\arraystretch}{1.0}
\setlength{\LTcapwidth}{\linewidth}
\setcounter{LTchunksize}{200}
\setlength{\LTpre}{4pt}\setlength{\LTpost}{4pt}
\setlength{\aboverulesep}{1pt}\setlength{\belowrulesep}{1pt}
\begin{longtable}{@{}c@{\hspace{8pt}}cccc@{\hspace{11pt}}cccc@{\hspace{11pt}}cccc@{}}
\caption{Per-seed test AUCs from retraining with the validation-selected hyperparameters used in Table~\ref{tab:training-variability}.}
\label{tab:repeat15-per-seed}\\
\toprule
 & \multicolumn{4}{c}{\textbf{Teacher-guided MTL}} & \multicolumn{4}{c}{\textbf{Traditional MTL}} & \multicolumn{4}{c}{\textbf{STL}} \\
\cmidrule(lr){2-5}\cmidrule(lr){6-9}\cmidrule(lr){10-13}
 & \makebox[0pt][c]{\scriptsize ShotAUC} & \multicolumn{3}{c}{\scriptsize WindowAUC} & \makebox[0pt][c]{\scriptsize ShotAUC} & \multicolumn{3}{c}{\scriptsize WindowAUC} & \makebox[0pt][c]{\scriptsize ShotAUC} & \multicolumn{3}{c}{\scriptsize WindowAUC} \\
\cmidrule(lr){2-2}\cmidrule(lr){3-5}\cmidrule(lr){6-6}\cmidrule(lr){7-9}\cmidrule(lr){10-10}\cmidrule(lr){11-13}
Seed ID & DP & ELM & \makebox[2.8em][c]{MARFE} & H/L & DP & ELM & \makebox[2.8em][c]{MARFE} & H/L & DP & ELM & \makebox[2.8em][c]{MARFE} & H/L \\
\midrule
\endfirsthead
\multicolumn{13}{@{}l@{}}{\tablename~\thetable\ (continued)}\\[3pt]
\toprule
 & \multicolumn{4}{c}{\textbf{Teacher-guided MTL}} & \multicolumn{4}{c}{\textbf{Traditional MTL}} & \multicolumn{4}{c}{\textbf{STL}} \\
\cmidrule(lr){2-5}\cmidrule(lr){6-9}\cmidrule(lr){10-13}
 & \makebox[0pt][c]{\scriptsize ShotAUC} & \multicolumn{3}{c}{\scriptsize WindowAUC} & \makebox[0pt][c]{\scriptsize ShotAUC} & \multicolumn{3}{c}{\scriptsize WindowAUC} & \makebox[0pt][c]{\scriptsize ShotAUC} & \multicolumn{3}{c}{\scriptsize WindowAUC} \\
\cmidrule(lr){2-2}\cmidrule(lr){3-5}\cmidrule(lr){6-6}\cmidrule(lr){7-9}\cmidrule(lr){10-10}\cmidrule(lr){11-13}
Seed ID & DP & ELM & \makebox[2.8em][c]{MARFE} & H/L & DP & ELM & \makebox[2.8em][c]{MARFE} & H/L & DP & ELM & \makebox[2.8em][c]{MARFE} & H/L \\
\midrule
\endhead
\midrule
\multicolumn{13}{r}{\footnotesize Continued on next page}\\
\endfoot
\bottomrule
\noalign{\vskip4pt}
\multicolumn{13}{@{}l@{}}{\begin{minipage}{\linewidth}\footnotesize\raggedright
Mean and sample SD use unrounded AUCs.
\end{minipage}}\\
\noalign{\vskip0pt}
\endlastfoot
1 & 0.9752 & 0.9938 & 0.9938 & 0.9977 & 0.9635 & 0.9928 & 0.9858 & 0.9979 & 0.9698 & 0.9923 & 0.9907 & 0.9909 \\
2 & 0.9724 & 0.9942 & 0.9936 & 0.9973 & 0.9602 & 0.9903 & 0.9887 & 0.9986 & 0.9673 & 0.9923 & 0.9910 & 0.9965 \\
3 & 0.9726 & 0.9929 & 0.9944 & 0.9975 & 0.9626 & 0.9895 & 0.9843 & 0.9986 & 0.9695 & 0.9935 & 0.9926 & 0.9942 \\
4 & 0.9712 & 0.9936 & 0.9939 & 0.9979 & 0.9669 & 0.9914 & 0.9912 & 0.9966 & 0.9656 & 0.9919 & 0.9903 & 0.9885 \\
5 & 0.9745 & 0.9938 & 0.9943 & 0.9983 & 0.9652 & 0.9926 & 0.9900 & 0.9985 & 0.9685 & 0.9923 & 0.9909 & 0.9966 \\
6 & 0.9710 & 0.9939 & 0.9941 & 0.9972 & 0.9648 & 0.9885 & 0.9893 & 0.9987 & 0.9657 & 0.9921 & 0.9910 & 0.9968 \\
7 & 0.9749 & 0.9936 & 0.9931 & 0.9980 & 0.9636 & 0.9933 & 0.9907 & 0.9988 & 0.9696 & 0.9937 & 0.9906 & 0.9913 \\
8 & 0.9754 & 0.9946 & 0.9950 & 0.9978 & 0.9668 & 0.9942 & 0.9893 & 0.9991 & 0.9670 & 0.9944 & 0.9920 & 0.9957 \\
9 & 0.9757 & 0.9929 & 0.9947 & 0.9978 & 0.9632 & 0.9898 & 0.9890 & 0.9975 & 0.9668 & 0.9911 & 0.9905 & 0.9723 \\
10 & 0.9725 & 0.9930 & 0.9940 & 0.9977 & 0.9664 & 0.9915 & 0.9890 & 0.9987 & 0.9692 & 0.9914 & 0.9899 & 0.9962 \\
11 & 0.9746 & 0.9917 & 0.9951 & 0.9979 & 0.9669 & 0.9930 & 0.9887 & 0.9986 & 0.9650 & 0.9921 & 0.9903 & 0.9956 \\
12 & 0.9721 & 0.9939 & 0.9941 & 0.9976 & 0.9647 & 0.9900 & 0.9878 & 0.9971 & 0.9660 & 0.9926 & 0.9907 & 0.9967 \\
13 & 0.9720 & 0.9931 & 0.9949 & 0.9982 & 0.9638 & 0.9904 & 0.9872 & 0.9983 & 0.9694 & 0.9924 & 0.9912 & 0.9884 \\
14 & 0.9696 & 0.9942 & 0.9938 & 0.9976 & 0.9642 & 0.9918 & 0.9877 & 0.9985 & 0.9676 & 0.9910 & 0.9910 & 0.9949 \\
15 & 0.9743 & 0.9922 & 0.9944 & 0.9981 & 0.9678 & 0.9910 & 0.9828 & 0.9985 & 0.9646 & 0.9958 & 0.9912 & 0.9823 \\
\midrule
\textbf{Mean} & 0.9732 & 0.9934 & 0.9942 & 0.9978 & 0.9647 & 0.9913 & 0.9881 & 0.9983 & 0.9674 & 0.9926 & 0.9909 & 0.9918 \\*
\textbf{SD} & 0.0019 & 0.0008 & 0.0006 & 0.0003 & 0.0020 & 0.0016 & 0.0023 & 0.0007 & 0.0018 & 0.0013 & 0.0007 & 0.0068 \\
\end{longtable}
\endgroup

\endgroup

\FloatBarrier
\section{Disruption warning times}
\label{app:warning-time-distributions}
Warning time (WT) is the interval from the first alarm to disruption, with no alarm assigned zero.
Each model's median WT uses the same 488 disruptive test shots at its test-set ROC threshold with shot-level FPR closest to 0.10; all observed alarm lead times are retained.
WT is reported descriptively and is not used to select STL baseline models, optimize MTL hyperparameters, select checkpoints, or assess negative transfer.
The DP performance metric used in hyperparameter optimization is validation ShotAUC.

\begingroup
\small
\captionsetup{font=normalsize}

\renewcommand{\arraystretch}{1.2}\setlength{\tabcolsep}{12pt}
\setlength{\LTcapwidth}{\linewidth}
\setcounter{LTchunksize}{200}
\setlength{\LTpre}{4pt}\setlength{\LTpost}{4pt}
\begin{longtable}{@{}lc@{}}
\caption{Test warning times from fifteen-seed retraining with the validation-selected hyperparameters used in Table~\ref{tab:training-variability}.}
\label{tab:warning-time-summary}\\
\toprule
\textbf{Method} & \textbf{Median WT (ms)} \\
\midrule
\endfirsthead
\multicolumn{2}{@{}l@{}}{\tablename~\thetable\ (continued)}\\[3pt]
\toprule
\textbf{Method} & \textbf{Median WT (ms)} \\
\midrule
\endhead
\midrule
\multicolumn{2}{r}{\footnotesize Continued on next page}\\
\endfoot
\bottomrule
\noalign{\vskip4pt}
\multicolumn{2}{@{}l@{}}{\begin{minipage}{\linewidth}\footnotesize\raggedright
Values are mean $\pm$ sample SD across the fifteen model-wise median WTs.
Teacher-guided MTL has a longer median WT than traditional MTL in 13 of 15 same-seed comparisons.
\end{minipage}}\\
\noalign{\vskip0pt}
\endlastfoot
Teacher-guided MTL & $231.6 \pm 11.6$ \\
Traditional MTL & $212.9 \pm 24.8$ \\
STL (DP) & $207.9 \pm 12.3$ \\
\end{longtable}
\endgroup

Table~\ref{tab:warning-time-seed1} reports the percentages of disruptive shots with longer, equal, or shorter WT in all three pairwise comparisons of the methods in Table~\ref{tab:warning-time-summary}, summarized across the same fifteen training seeds.
\begingroup
\small
\captionsetup{font=normalsize}
\renewcommand{\arraystretch}{1.2}\setlength{\tabcolsep}{4pt}
\setlength{\LTcapwidth}{\linewidth}
\setcounter{LTchunksize}{200}
\setlength{\LTpre}{4pt}\setlength{\LTpost}{4pt}
\begin{longtable}{@{}lccc@{}}
\caption{Per-shot WT changes across fifteen training seeds for the methods in Table~\ref{tab:warning-time-summary}.}
\label{tab:warning-time-seed1}\\
\toprule
& \multicolumn{3}{c}{\textbf{Fraction of disruptive shots (\%)}} \\
\cmidrule(l){2-4}
\textbf{$\Delta\mathrm{WT}$ definition} & \textbf{Longer WT} & \textbf{Equal WT} & \textbf{Shorter WT} \\
\midrule
\endfirsthead
\multicolumn{4}{@{}l@{}}{\tablename~\thetable\ (continued)}\\[3pt]
\toprule
& \multicolumn{3}{c}{\textbf{Fraction of disruptive shots (\%)}} \\
\cmidrule(l){2-4}
\textbf{$\Delta\mathrm{WT}$ definition} & \textbf{Longer WT} & \textbf{Equal WT} & \textbf{Shorter WT} \\
\midrule
\endhead
\midrule
\multicolumn{4}{r}{\footnotesize Continued on next page}\\
\endfoot
\bottomrule
\noalign{\vskip4pt}
\multicolumn{4}{@{}l@{}}{\begin{minipage}{\linewidth}\footnotesize\raggedright
Values are mean $\pm$ sample SD of the per-seed percentages across the fifteen training seeds used in Table~\ref{tab:warning-time-summary}.
Each seed uses the same 488 disruptive test shots, pairing models trained with the same seed by shot ID.
$\Delta\mathrm{WT}$ is the first model's WT minus the second's; equal WT includes pairs with no alarm from either model.
Percentages may not sum to 100 because of rounding.
\end{minipage}}\\
\noalign{\vskip0pt}
\endlastfoot
Teacher-guided MTL $-$ Traditional MTL & $61.1 \pm 6.7$ & $8.0 \pm 1.7$ & $30.9 \pm 6.0$ \\
Teacher-guided MTL $-$ STL (DP) & $52.8 \pm 5.1$ & $11.4 \pm 2.0$ & $35.7 \pm 4.4$ \\
Traditional MTL $-$ STL (DP) & $35.9 \pm 7.9$ & $9.8 \pm 1.9$ & $54.3 \pm 8.0$ \\
\end{longtable}
\endgroup

\clearpage
\begingroup\emergencystretch=2em
\section{Per-seed ablation results}\label{app:ablation15}
\begingroup
\font\AblationMetricFont=cmr8 at 7.6pt
\fontsize{9.5}{11.4}\selectfont
\captionsetup{font=normalsize}
\setlength{\tabcolsep}{2.5pt}\renewcommand{\arraystretch}{1.05}
\setlength{\LTcapwidth}{\linewidth}
\setcounter{LTchunksize}{200}
\setlength{\LTpre}{4pt}\setlength{\LTpost}{4pt}
\newlength{\AblationSeedColWidth}
\setlength{\AblationSeedColWidth}{\dimexpr(\linewidth-66pt)/8\relax}
\begin{longtable}{@{}>{\centering\arraybackslash}p{22pt}@{\hspace{4pt}}*{4}{>{\centering\arraybackslash}p{\AblationSeedColWidth}}@{\hspace{10pt}}*{4}{>{\centering\arraybackslash}p{\AblationSeedColWidth}}@{}}
\caption{Per-seed test AUCs for the four ablation conditions in Table~\ref{tab:component-effects}.}
\label{tab:ablation15-per-seed}\\
\toprule
\endfirsthead
\multicolumn{9}{@{}l@{}}{\tablename~\thetable\ (continued)}\\[3pt]
\toprule
\endhead
\midrule
\noalign{\nobreak}
\multicolumn{9}{r}{\footnotesize Continued on next page}\\
\endfoot
\bottomrule
\noalign{\vskip4pt}
\multicolumn{9}{@{}l@{}}{\begin{minipage}{\linewidth}\fontsize{9.5}{11.4}\selectfont\raggedright
The four conditions are paired by training seed.
Mean and sample SD are calculated from unrounded AUCs.
\end{minipage}}\\
\noalign{\vskip0pt}
\endlastfoot
 & \multicolumn{4}{c}{\textbf{\shortstack{Teacher-guided MTL\\(full)}}} & \multicolumn{4}{c}{\textbf{\shortstack{Without DP peak-disruptivity\\window weighting}}} \\*
\cmidrule(lr){2-5}\cmidrule(lr){6-9}
\noalign{\nobreak}
\makebox[0pt][c]{Seed ID} & \makebox[0pt][c]{\shortstack{DP\\{\AblationMetricFont ShotAUC}}} & \makebox[0pt][c]{\shortstack{ELM\\{\AblationMetricFont WindowAUC}}} & \makebox[0pt][c]{\shortstack{MARFE\\{\AblationMetricFont WindowAUC}}} & \makebox[0pt][c]{\shortstack{H/L\\{\AblationMetricFont WindowAUC}}} & \makebox[0pt][c]{\shortstack{DP\\{\AblationMetricFont ShotAUC}}} & \makebox[0pt][c]{\shortstack{ELM\\{\AblationMetricFont WindowAUC}}} & \makebox[0pt][c]{\shortstack{MARFE\\{\AblationMetricFont WindowAUC}}} & \makebox[0pt][c]{\shortstack{H/L\\{\AblationMetricFont WindowAUC}}} \\*
\midrule
\noalign{\nobreak}
1 & 0.9752 & 0.9938 & 0.9938 & 0.9977 & 0.9706 & 0.9933 & 0.9936 & 0.9975 \\*
2 & 0.9724 & 0.9942 & 0.9936 & 0.9973 & 0.9703 & 0.9944 & 0.9940 & 0.9976 \\*
3 & 0.9726 & 0.9929 & 0.9944 & 0.9975 & 0.9718 & 0.9933 & 0.9952 & 0.9975 \\*
4 & 0.9712 & 0.9936 & 0.9939 & 0.9979 & 0.9700 & 0.9938 & 0.9943 & 0.9976 \\*
5 & 0.9745 & 0.9938 & 0.9943 & 0.9983 & 0.9722 & 0.9945 & 0.9939 & 0.9981 \\*
6 & 0.9710 & 0.9939 & 0.9941 & 0.9972 & 0.9690 & 0.9934 & 0.9952 & 0.9983 \\*
7 & 0.9749 & 0.9936 & 0.9931 & 0.9980 & 0.9696 & 0.9928 & 0.9937 & 0.9975 \\*
8 & 0.9754 & 0.9946 & 0.9950 & 0.9978 & 0.9721 & 0.9922 & 0.9951 & 0.9982 \\*
9 & 0.9757 & 0.9929 & 0.9947 & 0.9978 & 0.9727 & 0.9918 & 0.9938 & 0.9983 \\*
10 & 0.9725 & 0.9930 & 0.9940 & 0.9977 & 0.9682 & 0.9933 & 0.9936 & 0.9978 \\*
11 & 0.9746 & 0.9917 & 0.9951 & 0.9979 & 0.9728 & 0.9928 & 0.9946 & 0.9976 \\*
12 & 0.9721 & 0.9939 & 0.9941 & 0.9976 & 0.9707 & 0.9919 & 0.9942 & 0.9977 \\*
13 & 0.9720 & 0.9931 & 0.9949 & 0.9982 & 0.9710 & 0.9935 & 0.9937 & 0.9979 \\*
14 & 0.9696 & 0.9942 & 0.9938 & 0.9976 & 0.9686 & 0.9932 & 0.9939 & 0.9975 \\*
15 & 0.9743 & 0.9922 & 0.9944 & 0.9981 & 0.9708 & 0.9927 & 0.9937 & 0.9982 \\*
\midrule
\noalign{\nobreak}
\makebox[0pt][c]{\textbf{Mean}} & 0.9732 & 0.9934 & 0.9942 & 0.9978 & 0.9707 & 0.9931 & 0.9942 & 0.9978 \\*
\makebox[0pt][c]{\textbf{SD}} & 0.0019 & 0.0008 & 0.0006 & 0.0003 & 0.0014 & 0.0008 & 0.0006 & 0.0003 \\
\noalign{\penalty0}
\midrule
\noalign{\nobreak}
 & \multicolumn{4}{c}{\textbf{\shortstack{Without soft-target\\distillation}}} & \multicolumn{4}{c}{\textbf{\shortstack{Without both teacher-guidance\\mechanisms}}} \\*
\cmidrule(lr){2-5}\cmidrule(lr){6-9}
\noalign{\nobreak}
\makebox[0pt][c]{Seed ID} & \makebox[0pt][c]{\shortstack{DP\\{\AblationMetricFont ShotAUC}}} & \makebox[0pt][c]{\shortstack{ELM\\{\AblationMetricFont WindowAUC}}} & \makebox[0pt][c]{\shortstack{MARFE\\{\AblationMetricFont WindowAUC}}} & \makebox[0pt][c]{\shortstack{H/L\\{\AblationMetricFont WindowAUC}}} & \makebox[0pt][c]{\shortstack{DP\\{\AblationMetricFont ShotAUC}}} & \makebox[0pt][c]{\shortstack{ELM\\{\AblationMetricFont WindowAUC}}} & \makebox[0pt][c]{\shortstack{MARFE\\{\AblationMetricFont WindowAUC}}} & \makebox[0pt][c]{\shortstack{H/L\\{\AblationMetricFont WindowAUC}}} \\*
\midrule
\noalign{\nobreak}
1 & 0.9705 & 0.9946 & 0.9843 & 0.9966 & 0.9581 & 0.9936 & 0.9865 & 0.9962 \\*
2 & 0.9703 & 0.9935 & 0.9875 & 0.9967 & 0.9598 & 0.9935 & 0.9878 & 0.9974 \\*
3 & 0.9697 & 0.9922 & 0.9864 & 0.9978 & 0.9586 & 0.9912 & 0.9886 & 0.9978 \\*
4 & 0.9704 & 0.9911 & 0.9859 & 0.9965 & 0.9639 & 0.9944 & 0.9846 & 0.9972 \\*
5 & 0.9732 & 0.9946 & 0.9875 & 0.9957 & 0.9651 & 0.9926 & 0.9896 & 0.9969 \\*
6 & 0.9705 & 0.9883 & 0.9896 & 0.9958 & 0.9606 & 0.9908 & 0.9906 & 0.9955 \\*
7 & 0.9707 & 0.9907 & 0.9804 & 0.9961 & 0.9586 & 0.9907 & 0.9878 & 0.9968 \\*
8 & 0.9701 & 0.9878 & 0.9827 & 0.9973 & 0.9585 & 0.9907 & 0.9803 & 0.9979 \\*
9 & 0.9713 & 0.9867 & 0.9876 & 0.9962 & 0.9611 & 0.9844 & 0.9850 & 0.9952 \\*
10 & 0.9720 & 0.9928 & 0.9850 & 0.9980 & 0.9558 & 0.9941 & 0.9803 & 0.9975 \\*
11 & 0.9736 & 0.9917 & 0.9798 & 0.9969 & 0.9489 & 0.9955 & 0.9876 & 0.9963 \\*
12 & 0.9752 & 0.9903 & 0.9690 & 0.9962 & 0.9590 & 0.9898 & 0.9822 & 0.9969 \\*
13 & 0.9602 & 0.9897 & 0.9827 & 0.9972 & 0.9635 & 0.9914 & 0.9903 & 0.9983 \\*
14 & 0.9726 & 0.9920 & 0.9833 & 0.9960 & 0.9561 & 0.9943 & 0.9891 & 0.9967 \\*
15 & 0.9713 & 0.9918 & 0.9858 & 0.9953 & 0.9524 & 0.9921 & 0.9883 & 0.9952 \\*
\midrule
\noalign{\nobreak}
\makebox[0pt][c]{\textbf{Mean}} & 0.9708 & 0.9912 & 0.9838 & 0.9966 & 0.9587 & 0.9919 & 0.9866 & 0.9968 \\*
\makebox[0pt][c]{\textbf{SD}} & 0.0033 & 0.0023 & 0.0049 & 0.0008 & 0.0043 & 0.0027 & 0.0034 & 0.0010 \\
\end{longtable}
\endgroup

\endgroup

\end{document}